\documentclass{article} %
\pdfoutput=1 

\usepackage{cite}
\usepackage{algorithm}
\usepackage{hyperref}
\usepackage[noend]{algpseudocode} 
\algnewcommand\algorithmicfunctionname{\textbf{function:}}
\algnewcommand\Functionname{\item[\algorithmicfunctionname]}
\algnewcommand\algorithmicinput{\textbf{input:}}
\algnewcommand\Input{\item[\algorithmicinput]}
\algnewcommand\algorithmicoutput{\textbf{output:}}
\algnewcommand\Output{\item[\algorithmicoutput]}
\usepackage{wrapfig}
\usepackage{fancybox}
\usepackage{dashbox}
\usepackage[english]{babel}
\usepackage{xcolor} %
\usepackage{amsmath}

\usepackage{units}
\usepackage{amsthm}
\usepackage{amssymb}
\usepackage{url}
\usepackage{balance}
\usepackage{multirow}
\usepackage{lscape}
\usepackage{rotating}
\usepackage[caption=false]{subfig}

\def\Id{\operatorname{Id}}

\usepackage{float}

\theoremstyle{plain}
\newtheorem{Satz}{Satz}[section]
\newtheorem{fact}[Satz]{Fact}

\begin{document}
\newboolean{showcomments}
\setboolean{showcomments}{false}
\ifthenelse{\boolean{showcomments}}
{ \newcommand{\mynote}[3]{
   \fbox{\bfseries\sffamily\scriptsize#1}
   {\small$\blacktriangleright$\textsf{\emph{\color{#3}{#2}}}$\blacktriangleleft$}}}
{ \newcommand{\mynote}[3]{}}
\newcommand{\ag}[1]{\mynote{Andres}{#1}{red}}
\newcommand{\todo}[1]{\mynote{TODO}{#1}{red}}
\newcommand{\jc}[1]{\mynote{Jeronimo}{#1}{blue}}
\newcommand{\se}[1]{\mynote{Sergio}{#1}{brown}}

\newcommand{\revi}[1]{{#1}} %

\newenvironment{allintypewriter}{\ttfamily}{\par}
\newcommand{\tn}[1]     {
  \textnormal{#1}
}
\newcommand{\set}[1]    {
  \ensuremath{
    \left\{ #1 \right\}
  }
}
\newcommand{\sset}[2] {
  \ensuremath{
    \left\{ \left.\,
      #1
    ~\right|~
      #2
    \,\right\}
  }
}

\title{Symmetry in Software Synthesis}

\author{Andr\'{e}s Goens\thanks{Authors are with the Center for Advancing Electronics Dresden (cfaed), Chair for Compiler Construction, TU Dresden, Germany }
\and
Sergio Siccha\thanks{Author is with the Lehr- und Forschungsgebiet Algebra, Lehrstuhl B f\"{u}r Mathematik, RWTH Aachen, Germany }
\and
Jeronimo Castrillon\footnotemark[1] 
}
\date{}
\maketitle
\begin{abstract}
With the surge of multi- and manycores, much research has focused on algorithms for mapping and scheduling on these complex platforms.
\revi{Large classes of these algorithms face scalability problems. This is why diverse} methods are commonly used for reducing the search space. %
While most such approaches leverage the inherent symmetry of architectures and applications, they do it in a problem-specific and intuitive way.
However, intuitive approaches become impractical with growing hardware complexity, like Network-on-Chip interconnect or heterogeneous cores.
\revi{In this paper, we present a formal framework that can determine the inherent symmetry of architectures and applications algorithmically and leverage these for problems in software synthesis.
Our approach is based on the mathematical theory of groups and a generalization called inverse semigroups. }
We evaluate our approach in two state-of-the-art mapping frameworks. %
Even for the platforms with a handful of cores of today and moderate-size benchmarks,
our approach \revi{consistently yields reductions of the overall execution time of algorithms, accelerating them by a factor up to $10$ in our experiments, or improving the quality of the results.} %

\end{abstract}

\section{Introduction}
Several multi-processor and many-core systems are available in the market today. %
Systems like embedded heterogeneous platforms, e.g.  TI Keystone II~\cite{keystone2_whitepaper} with multiple ARM and DSP cores, the homogeneous many-cores
provided by Adapteva~\cite{gwennap2011adapteva} or the ARM big.LITTLE platforms~\cite{biglittlewhitepaper} all show a trend, a trend that is clear:
Hardware architectures are becoming larger and more complex, for example with large heterogeneous systems with clusters of processing elements and Network-on-Chip (NoC) interconnect,
like is the case of the Kalray MPPA-256~\cite{kalraymppa256}.

\begin{figure}[h]
	\centering
	\includegraphics[width=0.60\textwidth]{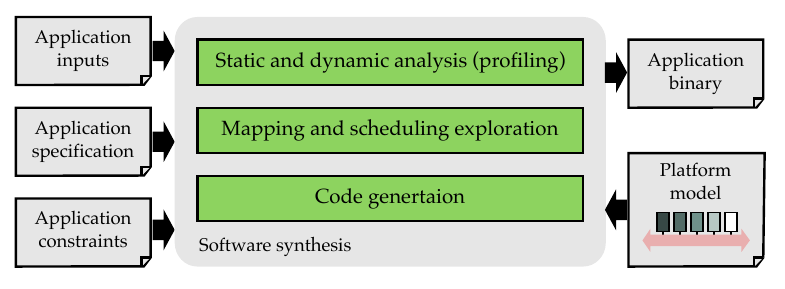}
	\caption{Abstract view of a software synthesis flow}
	\label{fig:synthesisflow}
\end{figure}

For these complex architectures, it is overly complicated for a designer to reason about parallel execution. 
Thus, the research community has invested a lot of effort in automatically deriving 
efficient implementations from abstract task-based specifications, 
particularly in the embedded domain~\cite{keinert2009systemcodesigner,thiele07acsd,castrillon2011trends}%
~(see Figure~\ref{fig:synthesisflow}).
The \emph{software synthesis} process includes defining the mapping of tasks and 
communication to resources, scheduling in case of resource sharing and the final 
code generation, among others. In the embedded domain, where predictability is usually as much an asset as performance,
the mapping is usually static and computed at compile time.
\revi{This class of approaches has a distinct advantage over distributed approaches since it leverages global knowledge of the architecture. As a result, it is possible to obtain significantly better and more reliable executions than with dynamic, distributed methods.
This is not without cost, however; this software synthesis approach requires a design-space exploration (DSE) with a huge space of design options, which usually grows exponentially with application and architecture sizes.}
Recent research has also focused on hybrid solutions, i.e., partially defined at compile time
with flexibility left for the runtime~\cite{quan2015hybrid,weichslgartner2014daarm}.

Due to the large size of the design-space, the DSE typically uses either custom heuristics or 
meta-heuristics for (multi-objective) optimization (e.g., evolutionary 
algorithms~\cite{deb2001multi}, 
particle swarm or ant colony optimization~\cite{dorigo2008ant}).
In this context, optimality refers to different objectives, like minimum execution time, energy consumption or the best load distribution to avoid thermal effects.
Examples are evolutionary algorithms in the programming 
frameworks Sesame~\cite{erbas_pimentel06} and 
DOL~\cite{thiele07acsd} and simpler iterative heuristics in SystemCoDesigner~\cite{keinert2009systemcodesigner}, in MAPS~\cite{castrillontii13} and in~\cite{Cheung07,casale2013design} among others.

\begin{figure}[t]
	\centering
	\includegraphics[width=0.65\textwidth]{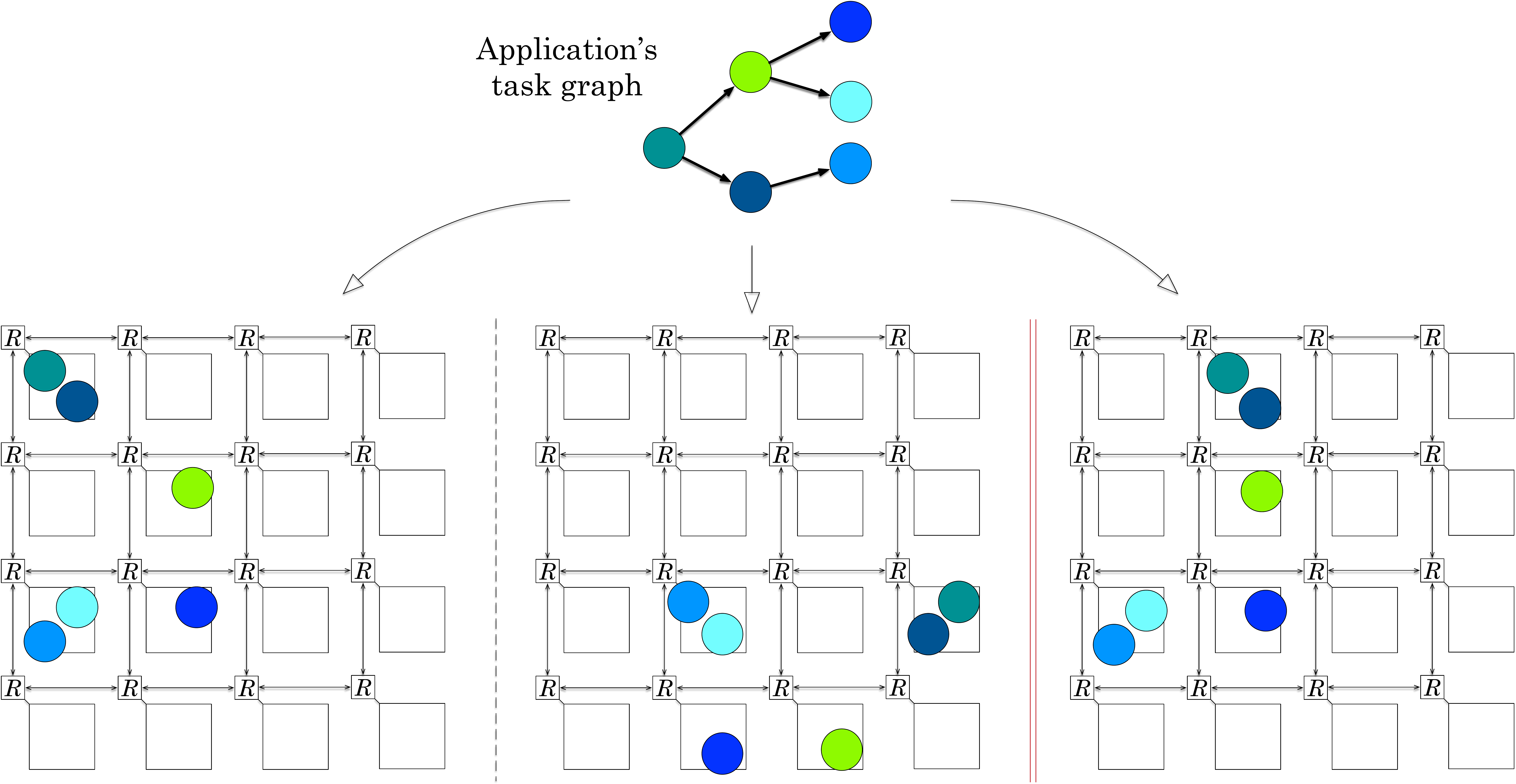}
	\caption{Illustration of symmetries. The two left mappings are equivalent, while the one on the right is not.}
	\label{fig:symmintro}
\end{figure}

To deal with the large design-space, it is very common for engineers and researchers to intuitively harness symmetry properties of the problem.
Approaches designed to work for homogeneous, bus-based architectures usually have a formulation that considers all cores to be identical. \ag{here we could cut a bit still}
When core heterogeneity is involved, only the type of the core is considered. In fact, the notions of homogeneous and heterogeneous architectures
are just approximations for describing symmetry in an architecture. For architectures with Network-on-Chip, the situation is more complex.
In those cases, most methods involve a hand-tailored reduction for the particular problem, like in ~\cite{varghese2015programming}, or pessimistic heuristics, e.g. in ~\cite{singh2013accelerating}.
To illustrate the principle, consider the example for a simple homogeneous many-core with a NoC interconnect shown in 
Figure~\ref{fig:symmintro}.
Due to the homogeneous structures of the platform, it is intuitive to see that the two leftmost mappings of tasks to processing nodes should lead to basically the same execution behavior, 
\revi{if we neglect effects like process variation or aging.}
Conversely, we can expect the third, rightmost mapping to have a different execution behavior, since the communication paths are different than in the first two. 
In the presence of heterogeneous resources, such an analysis becomes less.
Even more so with complex network topologies, in architectures with hierarchical structure, \revi{when optimizing for different objectives}, or when all these are combined.

In this paper, we introduce a general mathematical framework that allows reasoning about such general symmetries. 
It is based on the theory of groups and inverse semigroups, mathematical methods that enable us to define and quantify the symmetries.
With it, we can reduce the search space in an automated fashion, just as it is traditionally done by hand for specific problems. 
Additionally, the mathematical nature of the approach allows us to guarantee correctness and completeness of the described symmetry.
We demonstrate how our proposed methods can be applied to task-based programming models and to 
several kinds of architectures, including bus-based and NoC-based architectures.
We analyze algorithmic implementations to find symmetries and their complexity.
Finally, we integrate our approach into two state-of-the-art DSE frameworks,
showing a significant reduction in the number of candidates evaluated, even for today's
moderate-sized systems.

The rest of the paper is organized as follows. 
Sections~\ref{sec:inversesemigroups} and \ref{sec:actions} introduce the mathematical foundations for symmetry
and the notion of equivalent objects respectively,
while discussing their application in a software-synthesis context.
Section~\ref{sec:algorithmic} deals with the algorithmic realization,
while the approach is evaluated for two use-cases in Section~\ref{sec:eval}.
Finally, Sections~\ref{sec:related} and~\ref{sec:conclusions}
discuss related work and conclude the paper.

\section{Symmetries and Inverse Semigroups}
\label{sec:inversesemigroups}
Symmetry is a concept that is well-known intuitively, but hard to make precise without a rigorous treatment. 
An intuitive definition of symmetry is the quality of a system to be transformed, in a controlled manner, without changing its properties.
Modern hardware architectures exhibit a high degree of symmetry,
even in heterogeneous platforms which usually feature several processors of the same type. 
Similarly, applications can sometimes feature a significant amount of symmetry. 
For example, in some cases of data-level parallelism, when the mapping of two identical 
worker threads can be swapped without affecting performance.
Both application and architecture symmetries combined induce symmetry in the mapping of computation and communication to hardware resources.

In this section, we present a formal mathematical treatment of these intuitively motivated properties.
For this, we use the theory of groups, which is the conventional theory for describing symmetry in mathematics.
Considering communication in NoC-based architectures, we also motivate the need for a generalization using so-called inverse semigroups.

\subsection{Groups and symmetries}

\begin{wrapfigure}{r}{0.45\textwidth}
	\centering
	\includegraphics[width=0.45\textwidth]{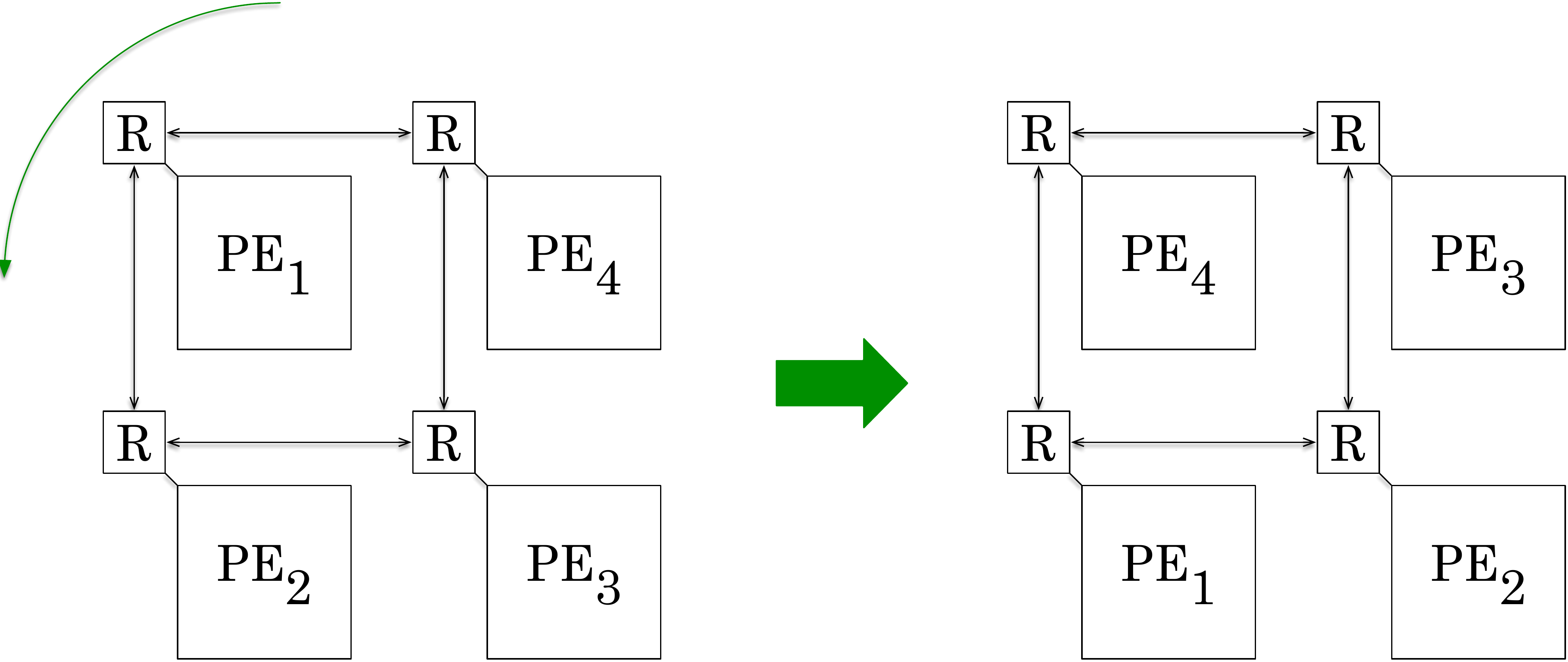}
	\caption{A 2x2 NoC architecture and its rotation by 90 degrees.}
   \label{fig:2x2rotation}
\end{wrapfigure}
Symmetry is a central concept in mathematics and has been studied extensively.
Most commonly, symmetry is described using the mathematical field of group theory.
\revi{Group theory studies symmetries of an object by analyzing the transformations which preserve the structure of the object. 
The set containing all these transformations is called a group.
More precisely, if we refer to the object as $\mathbb{O}$, said transformations are called the \emph{Group of Automorphisms} of $\mathbb{O}$.}
As an example, consider the architecture depicted in Figure~\ref{fig:2x2rotation}. 
It represents a homogeneous multi-processor system-on-chip (MPSoC) with four identical processing elements (PEs) connected by a two-by-two mesh NoC. Hence, $\mathbb{O} := \{ \tn{PE}_1 , \ldots, \tn{PE}_4 \}$.
\revi{Mathematically, we regard the transformations of this architecture as functions
$t \colon \mathbb{O} \rightarrow  \mathbb{O}$.
If we ignore communication, all four processing elements can be permuted arbitrarily, since they are identical.
This means that if we change the names of the PEs in any way, we still have the same architecture in principle.
However, if we consider communication, not every reordering of the PEs will respect the structure of the NoC.
In fact, there are precisely eight symmetry transformations which, when applied to this architecture, preserve its structure.}
The simplest one is doing nothing: this is usually called the trivial or identity transformation. 
A more interesting transformation consists of rotating \revi{the PEs} by $90^\circ$, as depicted in Figure~\ref{fig:2x2rotation}.
While the physical processors are not the same, they are all the same kind, and the intercommunication network also has the same structure.
Thus, we consider this rotated architecture on the right of Figure~\ref{fig:2x2rotation} to be equivalent to the one on the left.
Executing an application on this rotated architecture should yield the same results as on the original one, except for physical characteristics that cannot be predicted and accounted for at design time.

\begin{figure*}[t!]
	\subfloat[A reflection along the vertical axis\label{fig:2x2reflection}]{
     \begin{minipage}[b]{0.48\textwidth}
	\centering
	\includegraphics[width=1\textwidth]{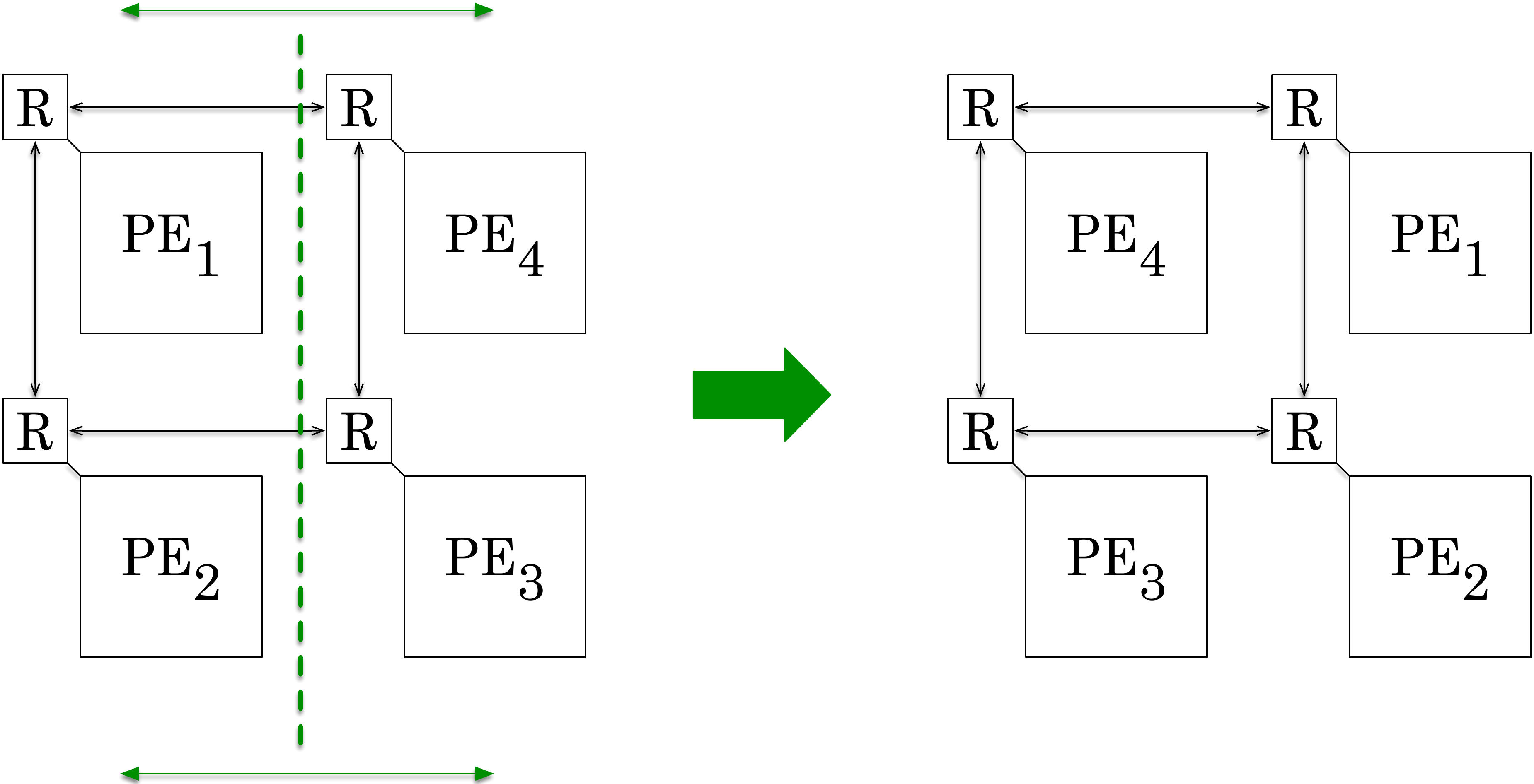}
		\end{minipage}
	}
	\hfill
	\centering
	\subfloat[A transformation that is not a symmetry\label{fig:2x2nosymmetry}]{
		\begin{minipage}[b]{0.48\textwidth}
	\centering
	\includegraphics[width=1\textwidth]{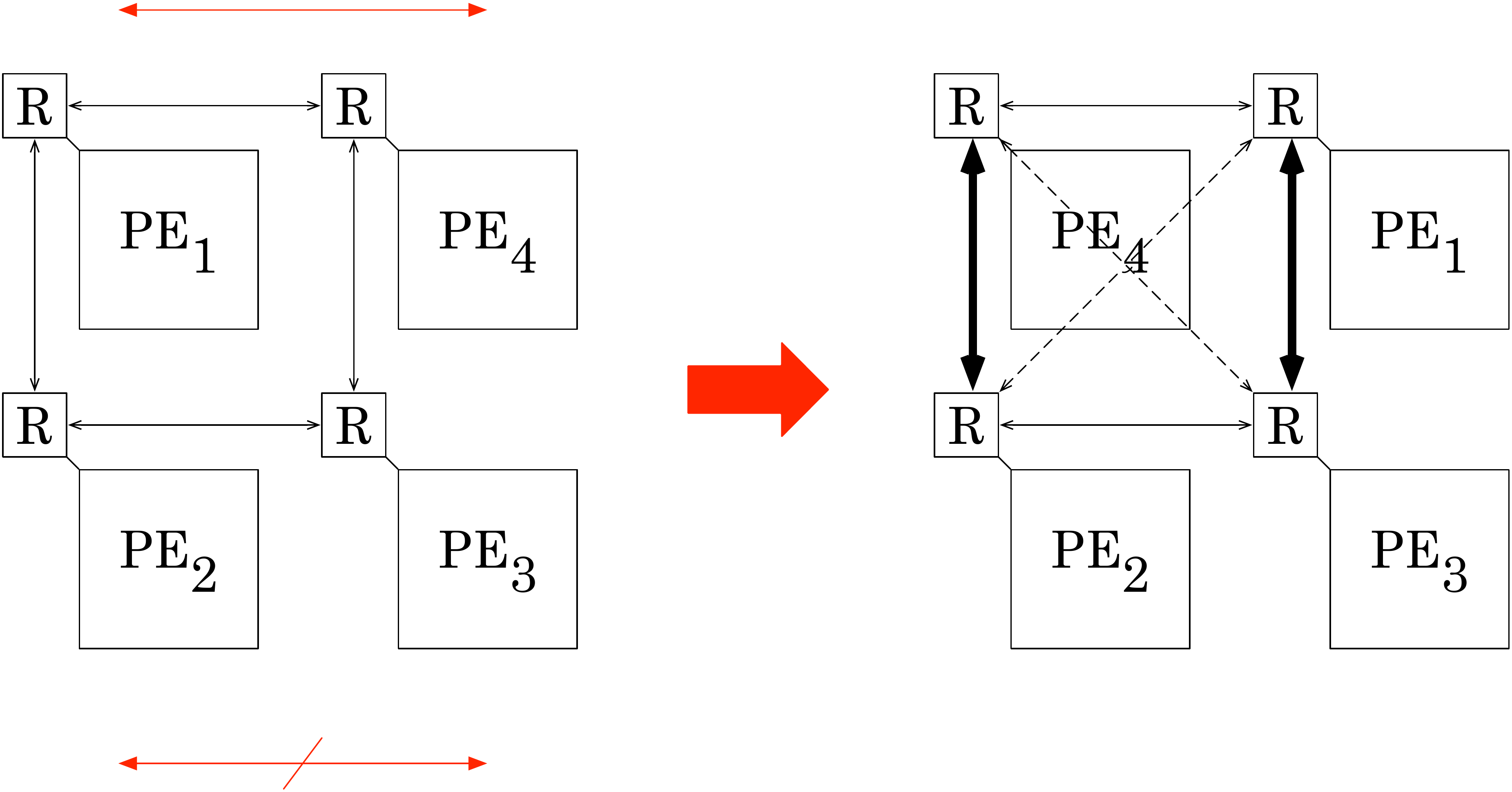}
		\end{minipage}
	}
	\caption{Symmetry transformations in a 2x2 NoC-based architecture}
	\label{fig:2x2symmetries}
\end{figure*}

We can repeat this rotation up to three times and get different transformations which preserve the structure of the whole architecture. 
After four rotations, the original architecture is obtained, which is equivalent to the identity transformation.
Reflection along the vertical axis is also a valid transformation, as depicted in Figure~\ref{fig:2x2reflection}.
In this case, the identity transformation results from applying reflection twice.
Furthermore, combining one fixed reflection with the three rotations of $90^\circ, 180^\circ, 270^\circ$ respectively yields three other reflections.
In total, there are eight possible transformations that preserve the structure of the architecture.
This group of transformations is well known as the \textsl{dihedral group} of order 8. 

As depicted in Figure~\ref{fig:2x2nosymmetry}, swapping the processors PE$_1$ and PE$_4$ is not a symmetry transformation when communication is considered. 
\revi{In the figure, the thicker communication lines were not in the architecture before the transformation, whereas the dotted lines were.}
In other words, the reason that this transformation is not a symmetry is that it does not preserve the communication costs between processors.
E.g., before applying the transformation, PE$_1$ is a direct neighbor of PE$_2$. Afterward, it is not. 

\revi{The transformations are thus captured in an abstract mathematical structure, a group, which we shall denote by $G$. 
The application of these transformations to the object $\mathbb O$ is called the \emph{action} of the group $G$ on $\mathbb O$,
whereas the elements of $\mathbb O$ are usually called \emph{points}.
Note that $G$ can also operate on another object $\mathbb O^\prime$ containing e.g.\ sub-architectures or mappings as points, instead of single PEs.
Groups are characterized by certain properties}\footnote{We omitted associativity in the axioms of a group since function composition is associative.}:
\begin{enumerate}
\item \label{prop:mult} Group elements, i.e., the symmetry transformations, can be combined and give rise to other symmetry transformations. 
In the case of the architecture, this means applying symmetry transformations one after another yields an additional symmetry of the architecture.
\item \label{prop:id} There is always a unique symmetry transformation that does nothing; it is called the identity and denoted by $\Id \in G$.
\item \label{prop:inv} Every symmetry transformation can be undone.
This means that for any symmetry transformation $t \in G$, there exists a unique symmetry transformation, $t^{-1}\in G$,
such that combining both yields the identity transformation: $tt^{-1} = t^{-1}t = \Id$.
In our case, for example, rotating by $90^\circ$ can be undone by rotating by $270^\circ$.
\end{enumerate}

Informal names for these properties are \emph{closedness under composition}, \emph{existence of a unique identity}, and \emph{existence of a unique inverse}.

\subsection{Symmetry as inverse semigroups of architectures}
\label{sec:groupsnotenough}

\begin{wrapfigure}{R}{.48\textwidth}
	\centering
	\includegraphics[width=.46\textwidth]{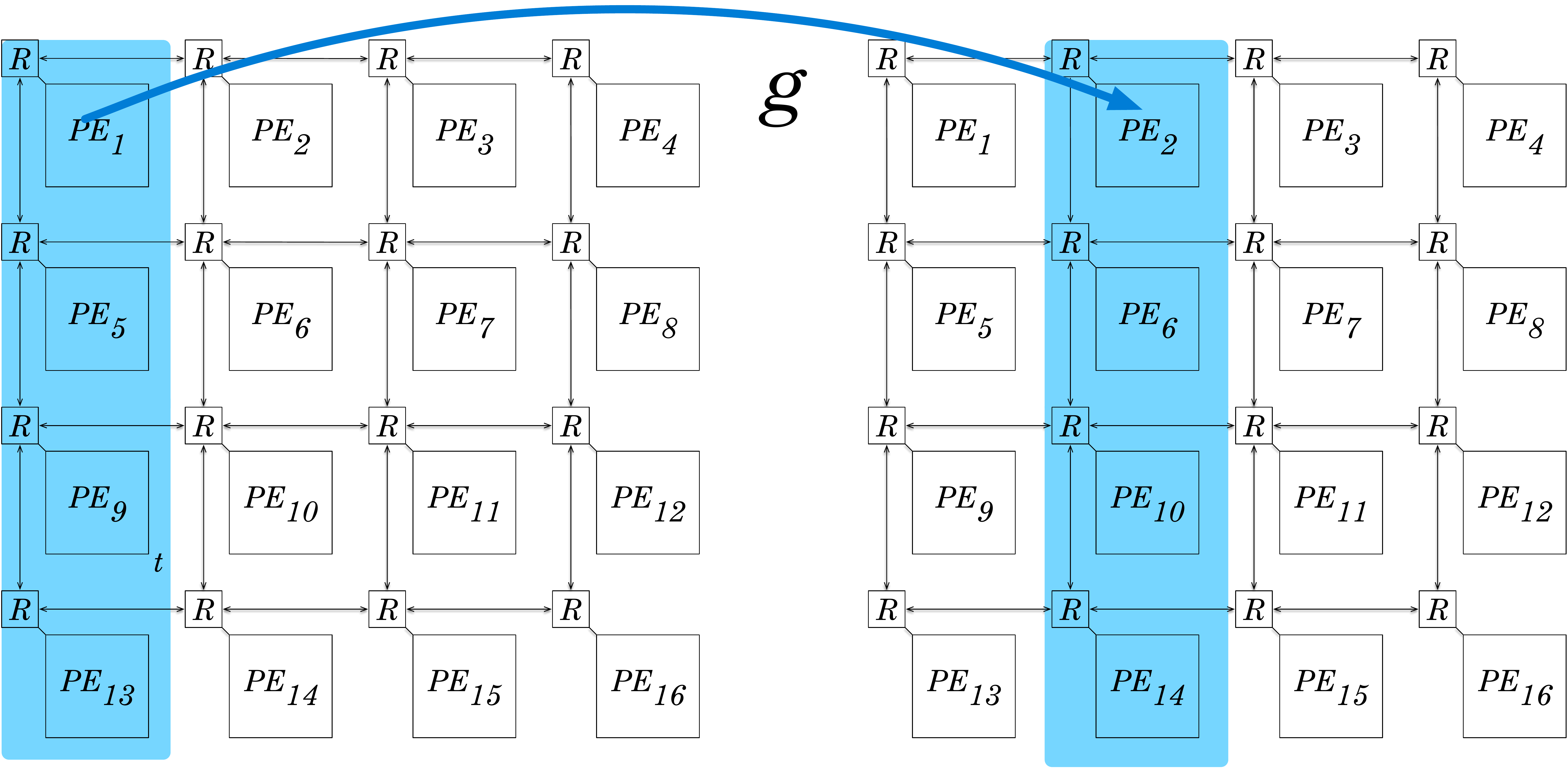}
	\caption{Partial functions define more symmetries in a 4x4 mesh.}
	\label{fig:4x4NoC_inv_semi}
\end{wrapfigure}

Consider the architecture depicted in Figure~\ref{fig:4x4NoC_inv_semi} as our object $\mathbb{O}$.
\revi{It represents a four-by-four NoC mesh with sixteen RISC PEs, each with a scratchpad memory. This architecture is inspired by the Parallela chip from Adapteva~\cite{gwennap2011adapteva}. However, this regular mesh structure 
can be found in several modern multicores, e.g.~\cite{openpiton,ramey2011_tilera}.}
For illustration purposes, we are not considering other factors like off-chip memories and peripherals in this example, 
\revi{but dealing with these is straightforward in this model, since they just change the symmetry group of the structure.}
The architecture from Figure~\ref{fig:4x4NoC_inv_semi} is similar to the architecture on the left of Figure~\ref{fig:2x2rotation}: \revi{they both have $n$-by-$n$ NoC topologies.}
In fact, by the formalism of symmetry groups, both architectures have the same symmetry.
There are only eight transformations that preserve the structure of the whole architecture and they are the same as those of the two-by-two mesh: the identity, three rotations and four reflections.  
The same is true for any such $n$-by-$n$ mesh NoC.%

\revi{However, intuitively, these are not the only symmetries of this architecture. To see this, compare the colored regions in the different depictions of the architecture in Figure~\ref{fig:4x4NoC_inv_semi}.
A computation using only the cores of the blue region to the left should have the same performance values as the one in the blue region to the right, with the transformation implied by the dark-blue arrow:}
\begin{align}
\label{eq:partial_4by4}\text{PE}_1 \mapsto \text{PE}_{2}, \text{PE}_{5} \mapsto \text{PE}_{6}, \text{PE}_9 \mapsto \text{PE}_{10}, \text{PE}_{13} \mapsto \text{PE}_{14}
\end{align}
The fundamental difference between this transformation and the aforementioned symmetries like rotations and reflections is, that it is only defined on a subset of the \revi{four-by-four} mesh and cannot be extended to a symmetry of the full mesh.
\revi{
In this context we will denote symmetry transformations of an object that are not defined on the whole object as partial symmetries.
Mathematically speaking, the partial symmetry transformations of this mesh do not form a group.
They do not satisfy property \ref{prop:inv} in its full rigor, as we shall see shortly.
Instead of a group, all partial symmetries of an object constitute what is called an inverse semigroup, which can be seen as a generalization of a group~\cite{lawson1998inversesemigroups}.
With inverse semigroups of symmetry, we do not consider functions $t \colon \mathbb{O} \rightarrow \mathbb{O}$. Instead we consider \emph{partial functions}, or more precisely, \emph{partial permutations}
$t' \colon \mathbb O~\rightarrow \mathbb O$, \emph{partial} meaning that the domain and co-domain (image) of $t'$ are only defined as subsets of $\mathbb O$.
Thus, $t'$ corresponds to a permutation $\mathcal{S}~\rightarrow~\mathcal{S}'$ from a sub-object (sub-set) $\mathcal{S} \subseteq \mathbb{O}$ to another (possibly different)
sub-object $\mathcal{S}' \subseteq \mathbb{O}$.
Due to this, when the co-domain of a transformation and the domain of another do not agree,
the composition of the transformations is only defined on the subset where they match (which can be empty). %

More precisely, we can equip the partial permutations with an extended composition rule such that they form an inverse semigroup:
if $f \colon X \to Y$ is a partial permutation
and $g \colon Y \to Z$ is a partial permutation, then their composition
$f \circ g \colon X \to Z$ is another partial permutation.
$f \circ g$ corresponds to a permutation from a subset of $X$ to a subset of $Z$ such that $(g \circ f)(x) = g(f(x))$ for all $x \in X$ where
$f(x)$ and $g(f(x))$ are defined, i.e., $x$ is in the domain of $f$ and $f(x)$ is in the domain of $g$.

For example, consider $X = Y = Z = \{1,\ldots,16\}$
and $f: 1 \mapsto 13, 5 \mapsto 9, 6 \mapsto 10, 7 \mapsto 11, 8 \mapsto 12, 12 \mapsto 8$ and $g: 1 \mapsto 2, 5 \mapsto 6, 9 \mapsto 10, 13 \mapsto 14$. We use this set instead of the PEs in the 4x4 NoC mesh for simplifying the notation. 
These two transformations are depicted in Figure~\ref{fig:partial_permutation}, $f$ as the green mapping  and $g$ as the blue one.
Then $gf = g \circ f$ is only defined for $\{1,5\}$, since only $f(1)$ and $f(5)$ are in $\{ 1,5,9,13 \}$, the domain of $g$.
Here $gf(1) = (g \circ f)(1) = g( f(1) ) = g( 13 ) = 14$, and similarly $gf(5) = g(f(5)) = g(9) = 10$, as can also be seen in Figure~\ref{fig:partial_permutation} by the orange mapping.

\begin{figure}
	\centering
	\includegraphics[width=.75\textwidth]{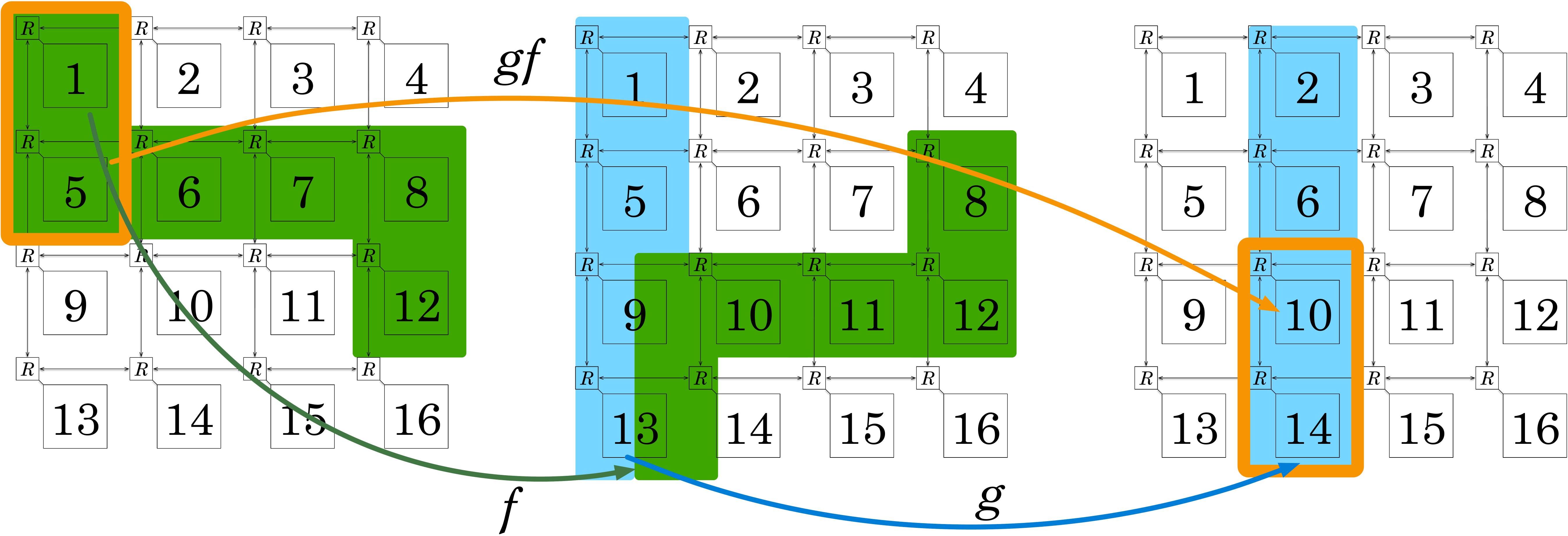}
	\caption{The product of partial permutations.}
	\label{fig:partial_permutation}
		\hspace{8mm}
\end{figure}

Groups have been studied much more in-depth than inverse semigroups have. With the extended composition rule we generalize several concepts from groups to use them for inverse semigroups.

To understand the difference between groups and inverse semigroups at a more formal level, consider again the transformation $t = g$, depicted in Figure~\ref{fig:partial_permutation}.
We stated earlier that general collections of partial permutations do not satisfy property (\ref{prop:inv})  in its full rigor.
Of course $g$ can be undone by a partial permutation $g^{-1}$. The resulting transformation $gg^{-1}$ is only defined on $\{ \tn{PE}_1 , \tn{PE}_5, \tn{PE}_9, \tn{PE}_{13} \}$ though.
We have $gg^{-1} \neq \Id_\mathbb{O}$, i.e.\ $gg^{-1}$ is not the identity transformation on the whole of $\mathbb{O}$,
which is not in line with property \ref{prop:inv}.
Instead, the product $gg^{-1} =: i $ is called a \emph{partial identity} on the subset $\{\text{PE}_1,\text{PE}_5,\text{PE}_9, \tn{PE}_{13}\}$,
i.e. $i = \Id\!\!\mid_{\{\text{PE}_1,\text{PE}_5,\text{PE}_9, \text{PE}_{13}\}}$, the restriction of the (global) identity to this subset.
In particular, it has the property that  $ii = i$ and elements with this property are usually called \emph{idempotents}.
In groups there is always exactly one idempotent, the identity.
In our inverse semigroups, the idempotents are exactly all partial identities.

A transformation $t$ in an inverse semigroup has an inverse element in a generalized sense, as illustrated above, which we will continue to denote as $t^{-1}$, by abuse of notation.
For our inverse semigroups, instead of the property \ref{prop:inv}, we get two new properties that are more general, albeit somewhat technical:
  \begin{enumerate}
  \item[(\ref{prop:id}')]Partial identities on subsets of the architecture commute: if we have two idempotents
      $i = \Id\!\!\mid_{I},f = \Id\!\!\mid_{F}$ for $I,F \subseteq \mathbb{O}$, then $if = fi = \Id\!\!\mid_{I \cap F}$.
  \item[(\ref{prop:inv}')]\label{prop:pseudoinv} For generalized inverse elements we do not have $tt^{-1} = \Id$, since $tt^{-1}$ is not defined on the whole set, unlike $\Id$. What does hold, however, are the equations
    \begin{align*} t t^{-1} t = t\text{ and } t^{-1} t t^{-1} = t^{-1}. \end{align*}
    This rule actually already implies that the product $tt^{-1}$ is an idempotent, as can be verified by a simple calculation.
  \end{enumerate}
  
In general, inverse semigroups need not satisfy property~(\ref{prop:id}') either, but those studied in this paper always do.
In technical terms, thus, our semigroups are also \emph{monoids}.
}

\revi{
The inverse semigroup of symmetries of the 4-by-4 mesh, for example, includes the partial function $t = g$ from Equation~\ref{eq:partial_4by4}. It yields an additional symmetry of the mesh which is not embraced by the group of architecture symmetries.
In general, inverse semigroups represent a richer set of symmetries by also considering symmetries between the substructures of an object. %

\subsection{Breaking symmetries}
In many occasions, the goal of software synthesis is not to optimize for latency, but e.g. power consumption or for avoiding thermal issues. Peripherials might be involved, or non-uniform access to main memory through the NoC.
We have not dealt with this in the examples above, but our methods are very capable of dealing with these scenarios.
Indeed, the concept of symmetry in terms of inverse semigroups, as introduced here, provides us with a \textbf{language to describe and leverage the inherent symmetries} of the hardware (and software).
The particular descriptions in this paper are mere examples of this.
If a perpherial can only be accessed through a controller on one end of the NoC, then the symmetries studied become partial symmetries in a larger inverse semigroup.
It takes said peripherial into account and thus has less global symmetries. %
}

\section{Equivalence of Mappings}
\label{sec:actions}
The formalism of groups and inverse semigroups can describe the symmetries of architectures and applications. In order to leverage them, however, \revi{one} has to understand how this affects the mappings of software applications to hardware resources.
In this section, we explain how the formal concepts of symmetry can be used to define when two mappings of computation and communication to hardware resources are equivalent.

\subsection{Equivalence of mappings}
\label{sec:actionsonmappings}

As motivated intuitively in the introduction (cf. Figure~\ref{fig:symmintro}), different mappings of computation to hardware resources (PEs and communication) can be considered equivalent in many cases.
For a given metric (e.g. latency or energy consumption), if two mappings of computation to hardware should yield the same results, only one has to be evaluated.
Sometimes, on the other hand, we might explicitly want to use a different mapping, but retain the same performance guarantees: this might be interesting, for example, when optimizing thermal distribution within the chip, or to select an alternative mapping at run-time depending on the available platform resources.
Thus, these methods apply for a larger scale of the software synthesis process. They go beyond reducing the design-space for exploration.

\begin{figure*}[t!]
	\subfloat[The group orbit of $\{1,2,3,5,6,7\}$.\label{fig:orbit_group}]{
     \begin{minipage}[b]{0.41\textwidth}
	\centering
	\includegraphics[width=0.8\textwidth]{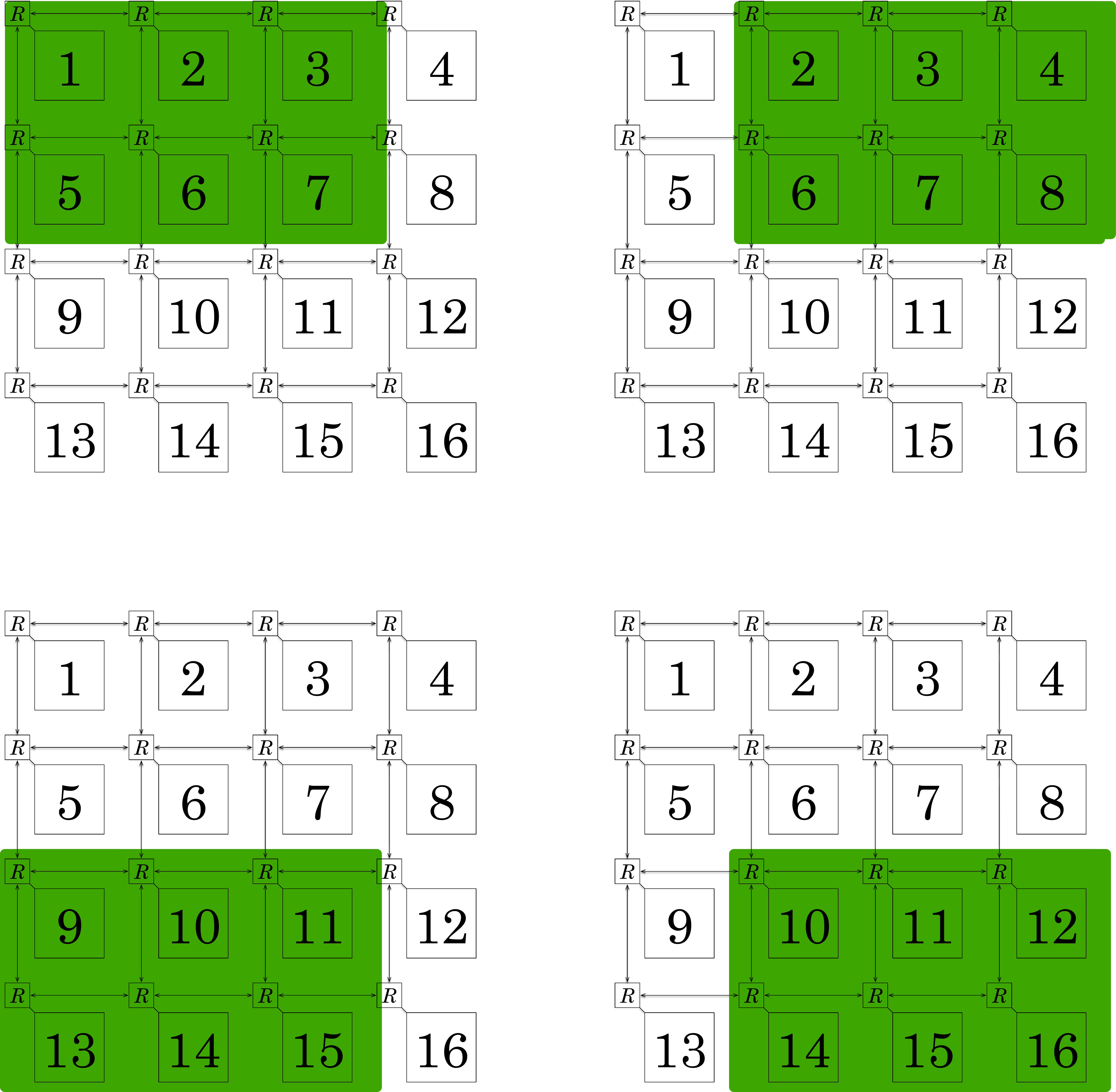}
		\end{minipage}
	}
	\hfill
	\centering
	\subfloat[Additional elements of the inverse semigroup orbit.\label{fig:orbit_semi}]{
		\begin{minipage}[b]{0.55\textwidth}
	\centering
	\includegraphics[width=0.54\textwidth]{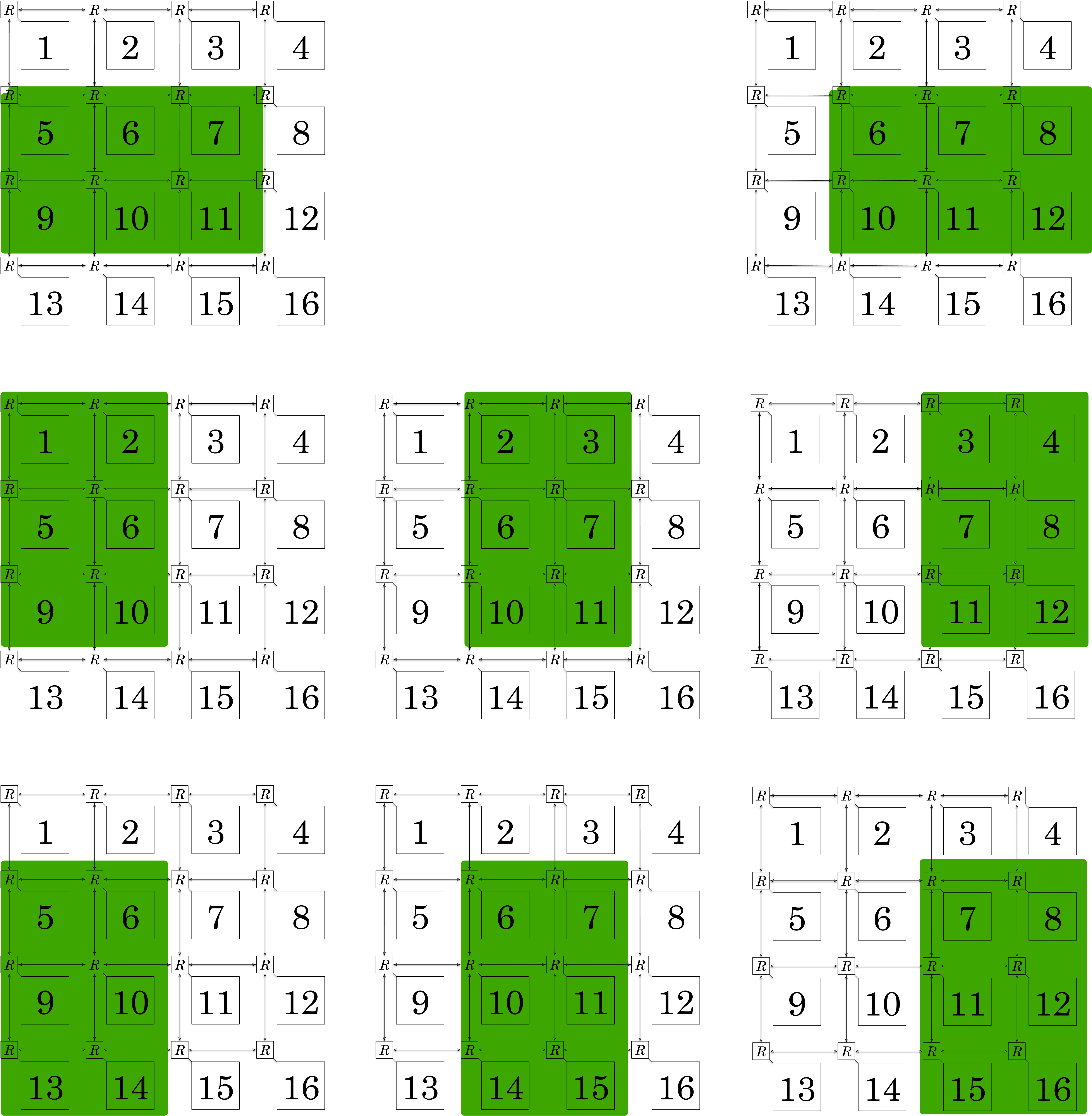}
		\end{minipage}
	}
	\caption{Examples of orbits on a 4x4 mesh NoC.}
	\label{fig:actionsm1}
\end{figure*}

We collect equivalent elements of the objects, mappings in this case, in sets. These sets are called \emph{orbits} of the \revi{group/inverse semigroup \emph{action}}\footnote{\revi{In the literature for inverse semigroups these are usually called strong orbits, or the strongly-connected components of the orbit graph, but we will use this simpler definition in this paper.}}.
To explain actions and orbits we refer again \revi{to the four-by-four mesh architecture (cf. Figure~\ref{fig:partial_permutation}).
As was mentioned before, if we consider the \textbf{group} $G$ of symmetries, it consists of the same eight transformations as the two-by-two mesh.}
A group action describes a way to apply a group element, or transformation, $t$ to a point $o \in \mathbb O$.
We denote the result of this application by $t \cdot m$.
Here, we choose the action $t \cdot o := t(o)$, i.e., simply apply the transformation function.
Given an action, we can then define the orbit of a point $o \in \mathbb O$ to be
\[
  G \cdot o := \{\, t( o ) \,|\, t \in G \,\}.
\]

\revi{
For example, consider the mesh itself as our object, $\mathbb O = \{ 1 \ldots 16 \}$, with our simplified the notation.
Consider then the orbit $G \cdot 1$ and let $t_1 = \varphi$ denote the anti-clockwise rotation by $90^\circ$. 
Then $13 \in G \cdot 1$, since $\varphi( 1 ) = 13$ and since $4 = \varphi(16) = \varphi(\varphi(13))$ and $\varphi(4) = 1$, we get 
$G \cdot 1 = \{1,13,16,4\}$, the set of all corner elements. Applying the other elements, i.e. the reflections, yields no additional elements in the orbit.

If we consider the inverse semigroup of symmetry $S$, since all PEs are equivalent, the orbit of $1$ under $S$ is the whole architecture, $S \cdot 1 = \mathbb{O}$.
If instead of a single PE, however, we consider the orbit of the subarchitecture $\{1,2,3,5,6,7\}$, then under the symmetry group, the orbit has four elements, namely
$\{2,3,4,6,7,8\}, \{9,10,11,13,14,15\}, \{10,11,12,14,15,16\}$, as depicted in Figure~\ref{fig:orbit_group}. In the case of the semigroup, on the other hand, another additional
8 elements are contained in the orbit, those depicted in Figure~\ref{fig:orbit_semi}.

 }

A more interesting example is the orbit of a mapping.
Let $T = \set{ T_1, \ldots, T_s }$ denote a set of tasks and $P = \set{ \tn{PE}_1, \ldots, \tn{PE}_r }$ a set of processing elements.
In the previous section, we mentioned that symmetries can act both on the architecture, i.e. on $P$, and also on the tasks $T$.
Let $H$ denote the group of symmetries of the tasks and $G$ the group of symmetries of the processors. 
As the object set, we define $\mathbb O$ to be the set of all task-to-processor mappings $m \colon T \rightarrow P$.
The symmetries described by these two groups are independent of each other for a mapping, and we can think of them as two groups acting separately. This is called the \emph{direct product} of $G$ and $H$, and is written as $G \times H$.

\begin{figure}
	\centering
	\includegraphics[width=.65\textwidth]{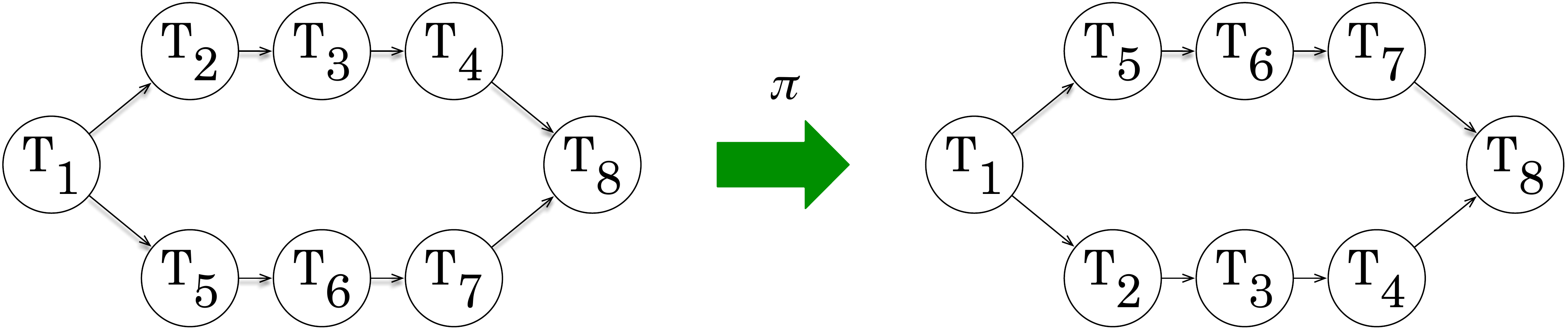}
	\caption{An example of an application: a stereo filter. Swapping the pipelines acting on the two channels, $\pi$, is a symmetry of this application.}
	\label{fig:audio_filter}
\end{figure}
\revi{
Consider as an example the application depicted in Figure~\ref{fig:audio_filter}. It describes an audio filter, where a source task $T_1$ separates the stereo input into two channels.
For each channel an inverse fourier transform ($T_2,T_5$), a filter on the frequency domain ($T_3,T_6$) and a fourier transform ($T_4,T_7$) are applied and gathered at a sink $T_8$.
This performs the same computation on the pipeline $T_2,T_3,T_4$ as in $T_5,T_6,T_7$, such that swapping both (complete) pipelines has no effect on the performance. We call this transformation on the application graph $\pi$,
see Figure~\ref{fig:audio_filter}.
Let us assume that we want to map this audio filter to the $r = 4$ processors of the architecture from Figure~\ref{fig:2x2rotation}.
To simplify the notation, we identify a mapping $m \colon T \rightarrow P$ with a tuple of the form
$\left(
  m(T_1), \ldots, m(T_s)
\right)$.
This way, the mapping
$
m_1 \colon
  T_1 \rightarrow \tn{PE}_2,
  T_2 \rightarrow \tn{PE}_3,
  T_3 \rightarrow \tn{PE}_3,
  T_4 \rightarrow \tn{PE}_3,
  T_5 \rightarrow \tn{PE}_4,
  T_6 \rightarrow \tn{PE}_4,
  T_7 \rightarrow \tn{PE}_4,
  T_8 \rightarrow \tn{PE}_1
$
can be written as
$m_1 = (\tn{PE}_2, \tn{PE}_3, \tn{PE}_3, \tn{PE}_3, \tn{PE}_4, \tn{PE}_4, \tn{PE}_4, \tn{PE}_1) = (2, 3, 3, 3, 4, 4, 4, 1)$,
for example, again simplifying the notation. %
\begin{figure*}[t!]
	\subfloat[The action of the architecture symmetry $\tau$.\label{fig:actiontau}]{
		\begin{minipage}[b]{0.48\textwidth}
	\centering
	\includegraphics[width=1\textwidth]{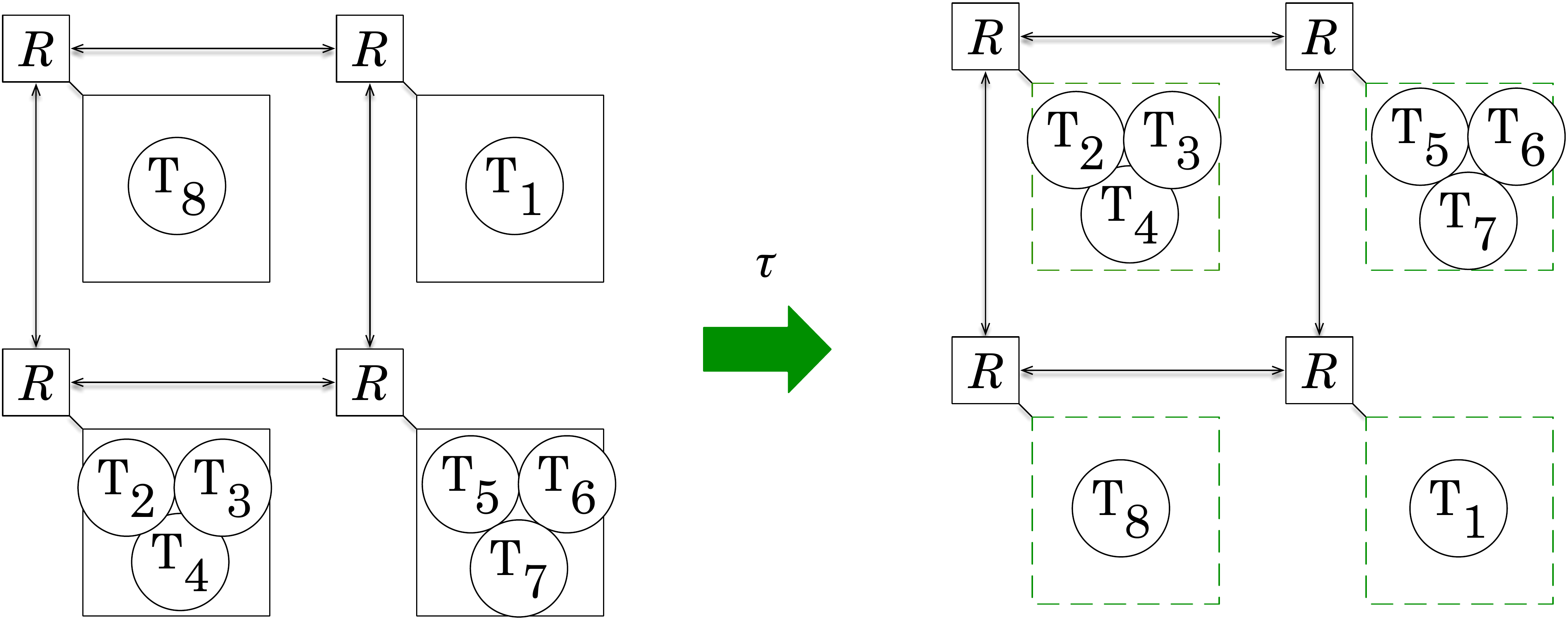}
		\end{minipage}
	}
	\hfill
	\centering
	\subfloat[The action of the task symmetry $\pi$.\label{fig:actionpi}]{
     \begin{minipage}[b]{0.48\textwidth}
	\centering
	\includegraphics[width=1\textwidth]{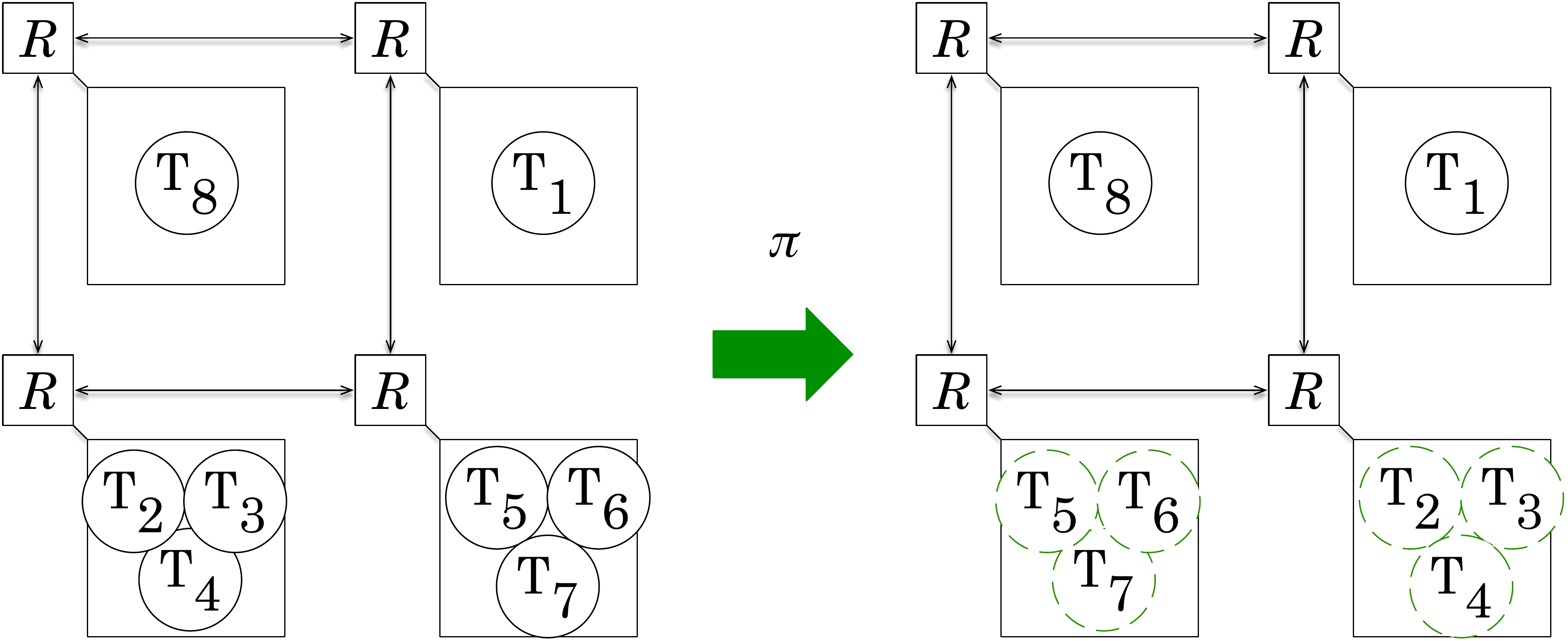}
		\end{minipage}
	}
	\caption{Actions of symmetries on the mapping $m_1$. Dotted lines denote the changes in the mapping.}
\end{figure*}

To understand the action, consider the symmetry of the architecture $\tau$, a reflection through the horizontal line (with respect to the depiction), which swaps $\tn{PE}_1$ and $\tn{PE}_3$ as well as $\tn{PE}_2$ and $\tn{PE}_4$. %
The action of a symmetry of the architecture, i.e. an element of $G$ like $\pi$ on $\mathbb O$ is straight-forward.
A permutation is just applied to each entry of a tuple $m$ individually.
Thus, as depicted in Figure~\ref{fig:actiontau} for $\tau \in G$ we have
\[
  \tau \cdot m_1 = ( \tau(2), \tau(3), \tau(3), \tau(3), \tau(4), \tau(4), \tau(4), \tau(1) ) = (4,1,1,1,2,2,2,3).
\]

A permutation of the tasks, in contrast to a symmetry of the architecture,
permutes the entries of a tuple horizontally, i.e. permutes the indices in the tuple.
For example, the permutation $\pi$ on the application (Figure~\ref{fig:audio_filter}) swapping the tasks $2,3,4$ and $5,6,7$  would swap the second, third and fourth entries of the tuple representing $m_1$ for its fifth, sixth and seventh entries respectively:
\[
  \pi \cdot m_1 = \pi \cdot (2, 3, 3, 3, 4, 4, 4, 1) = (2, 4, 4, 4, 3, 3, 3, 1) .
\]
This is illustrated in Figure~\ref{fig:actionpi}. Note that all three pairs of tasks have to be swapped simultaneously.
Using the action from just $\pi$, and the symmetry $\tau$, the orbit of $m_1$ would be
\begin{align*} \{(4, 1, 1, 1, 2, 2, 2, 3), (4, 2, 2, 2, 1, 1, 1, 3), (2, 4, 4, 4, 3, 3, 3, 1), (1, 3, 3, 3, 2, 2, 2, 4),\\
 (3, 1, 1, 1, 4, 4, 4, 2), (1, 2, 2, 2, 3, 3, 3, 4), (3, 4, 4, 4, 1, 1, 1, 2), (2, 3, 3, 3, 4, 4, 4, 1)\} \end{align*}
These mappings constitute the equivalence class of $m_1 = (2, 3, 3, 3, 4, 4, 4, 1)$ and should therefore behave equally in any execution.
}

\section{Algorithmic Considerations}
\label{sec:algorithmic}
Up to this point, we have introduced formal concepts to quantify and identify symmetries in software synthesis precisely.
These comprise groups, inverse semigroups, and orbits e.g.\ of a mapping or a sub-architecture.
\revi{%
In order to leverage these concepts, we design an algorithm that, for a given architecture, finds the inverse semigroup of its symmetries.
Then, we can use a standard algorithm to calculate the orbit of a given point at compile time.
Both, groups and inverse semigroups, can be very large, and it is infeasible to store all their elements explicitly.
The need for a compact representation %
in turn, has led to the development of data structures, which can be used to work with these compact representations.
}

The computer algebra system GAP~\cite{GAP4} implements many well-known algorithms in computational group theory.
Currently, the GAP community is working towards making these also available in its parallel fork HPC-GAP~\cite{HPC-GAP}.
In this paper, we use this computer-algebra system as a research vehicle, to assess the effectiveness of symmetry considerations in this field,
while we plan to use the parallel fork in future work.

This section describes the basic algorithmic concepts of our proposed methods,
discusses the overhead incurred by symmetry-related calculations and potential for domain-specific optimizations.

\subsection{A compact representation of groups and inverse semigroups}

For practical purposes it is prohibitive to store \revi{or iterate over} all elements of a group.
Instead, a group $G$ is usually given by a \emph{generating set} $S$, written $G = \langle S \rangle$.
We say the finite group $G$ is \emph{generated} by $S$, if $G$ consists of all words in the elements of $S$.
The advantage is that arbitrarily large groups can be written with small generating sets.
An important example is the so-called \emph{symmetric group}, written $S_n$, consisting of all permutations on $n$ points.
It has $n! = 1 \cdot \ldots \cdot n$ elements, but can always be generated by the following two elements:
a transposition $\tau \colon 1 \mapsto 2, 2 \mapsto 1, 3 \mapsto 3, \ldots, n \mapsto n$ which only swaps the points $1$ and $2$
and the $n$-cycle $\pi \colon 1 \mapsto 2 \mapsto 3 \mapsto \ldots \mapsto n-1 \mapsto n \mapsto 1$.
This representation is obviously very compact, since it uses only 2 permutations, each moving up to $n$ points, to represent the whole group of $n!$ many permutations.
It is a well-known fact that $S_n = \langle \tau, \pi \rangle$, i.e. any permutation on $n$ points can be written as a word in these two.
\revi{In general, there exist several generating sets for a given group or inverse semigroup which may vary in their sizes and algorithmic properties.
For a group, given by a generating set $S$ of permutations on $n$ points,
many problems can be solved in polynomial time w.r.t. $|S|$ and $n$.

\revi{%
For our purposes, the most fundamental of these problems is the \emph{membership test}.
While computing a group or an inverse semigroup of symmetries, we collect the symmetries already found
in a set $S$ and consider the $G = \langle S \rangle$.
Upon finding another symmetry $t$, we need to decide whether $t$ is a ``new'' symmetry and needs to be added to $S$,
or whether $t$ can be written as a word in the symmetries in $S$ collected so far.
Deciding whether $t \in \langle S \rangle$ is called the \emph{membership test}.
Being able to solve membership tests efficiently is crucial when working with a group or an inverse semigroup given by generators.
Usually, a data structure is computed which facilitates fast \emph{membership tests}.
An algorithm to solve this for groups is the \emph{Schreier-Sims algorithm}, first developed by C. Sims \cite{sims}.
It continues to be one of the main tools of computational group theory.

Since the inception of the Schreier-Sims algorithm, a tool-suite for working with arbitrary finite groups of permutations has been developed.
These methods and many more are described by Serres in his book \cite{seress}. 
State of the art implementations of these algorithms are available in GAP \cite{GAP4} and Magma \cite{magma},
and mostly run in Monte Carlo nearly-linear time $O^\sim( n \operatorname{log} n \, |S| )$.
}

}

\subsection{Computing the orbits}

\revi{%
To compute the orbit $G \cdot m$, for both groups and semigroups, we can employ generating sets. 
Starting with a given point $m$, 
the simplest algorithm applies all generators of the group $G$ to all points already encountered, repeatedly,
until no new points are found.
Then we can be sure that the full orbit has been calculated.
In general, for a group $G = \langle S \rangle$ and a point $m \in \mathbb O$, the orbit $G \cdot m$ can be computed with 
$\mathcal{O}( | G \cdot m | | S | )$ evaluations of the form $g \cdot m$~\cite{seress}.
In particular, the cardinality of the group is not relevant for the complexity of the algorithm, only the cardinalities of the orbit $G \cdot m$ and the generating set $S$ are relevant, which are almost always significantly smaller.
}
This is called the orbit algorithm.

\revi{%
In our case, the aim of using orbits of groups is only to identify whether two elements of the orbit are identical, for which we do not need to calculate the complete orbit.
Instead, if a unique representative of every orbit can always be chosen, we can compare the orbits of two elements by comparing their unique representatives.
This is sometimes called \emph{canonization} or finding a \emph{canonical representative}.
Such a canonical representative can be found when there is an ordering on the points in $\mathbb O$, %
in which case the canonical representative of an orbit $G \cdot m$ can be chosen to be its minimal element.
A special case of this (for the group $S_n$) yields the method employed in~\cite{thompson2013exploiting}.

To be able to calculate the orbit of a mapping, thus, we first need to compute a generating set of the automorphism group or semigroup.
However, we only have to do this once for every architecture. If we have a generating set, we can use it in any subsequent orbit calculation.
}

\subsection{Computing symmetry groups and inverse semigroups}

\begin{figure*}[t!]
	\subfloat[The topology graph of a $4 \times 4$ mesh NoC.]{
     \begin{minipage}[b]{0.45\textwidth}
	\centering
	\includegraphics[width=0.8\textwidth]{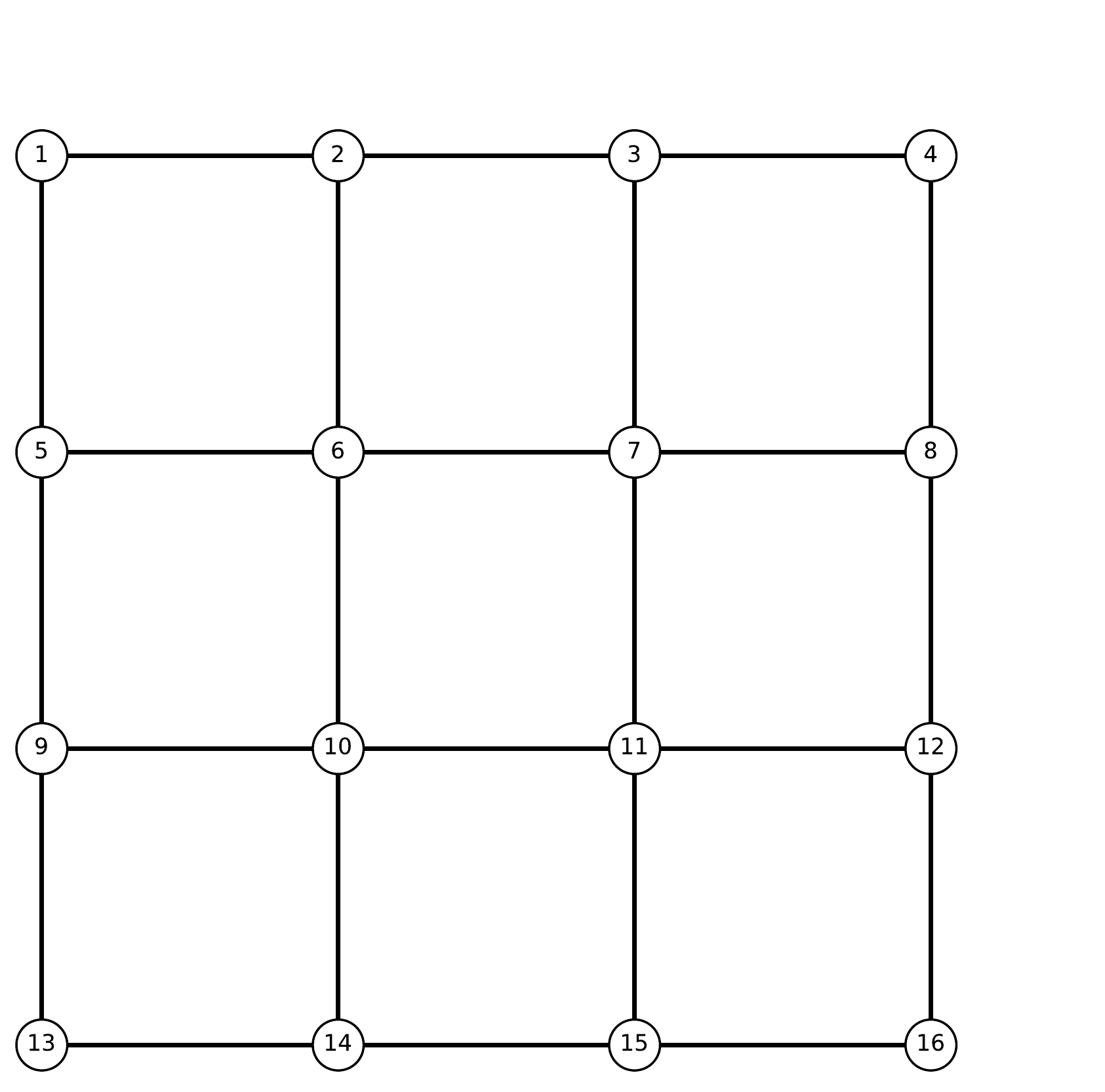}
		\end{minipage}
	}
	\centering
	\subfloat[The derived architecture graph. Edge colors represent the number of hops, darker is larger.]{
		\begin{minipage}[b]{0.45\textwidth}
	\centering
	\includegraphics[width=0.8\textwidth]{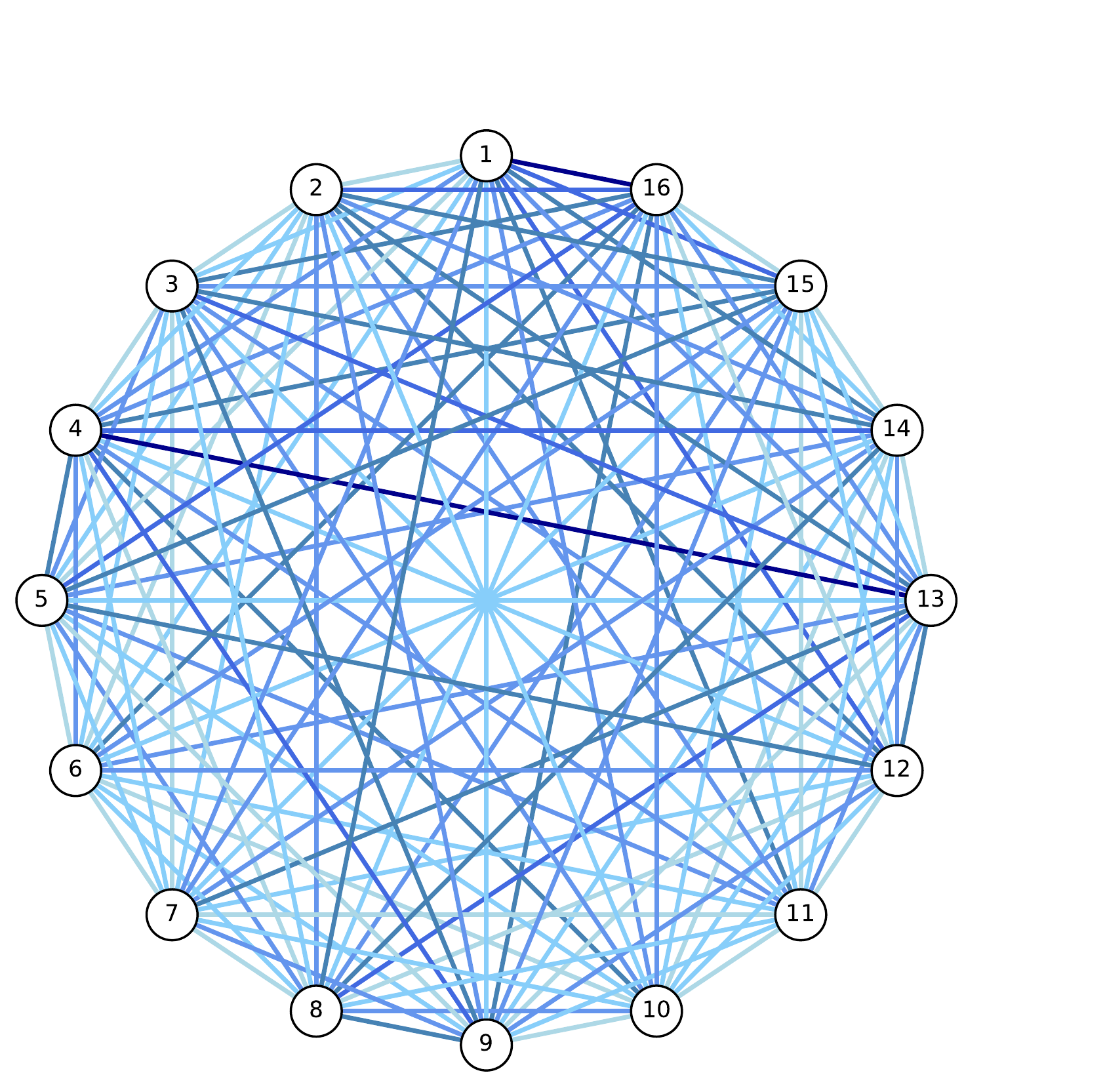}
		\end{minipage}
	}
	\caption{The architecture graph construction.}
	\label{fig:arch_graph}
\end{figure*}

Both, architectures and the relationships between tasks, can be modelled mathematically using graphs.
We model an architecture by a multigraph which has a node for every PE, labeled with the PE type. For every different communication resource between two PEs,
the graph has an edge annotated with that communication resource~\cite{castrillon2012}. This graph is sometimes called the \emph{architecture graph}. Figure~\ref{fig:arch_graph} illustrates this on the example of the graph representation of a $4 \times 4$ 
homogeneous NoC-mesh architecture. The annotations are not shown for readability. We construct this kind of graph easily from architecture models. However, there are no widespread standard formal models for this~\cite{goens_mcsoc16}, 
even though efforts have been made in this direction, like the SHIM standard from the Multicore Association~\cite{shim} or at a formal level~\cite{pelcat_models_of_architecture}.

On the application side, the situation is less straightforward. \revi{Most models of computation are based on a sequential, shared memory model, which means they are not well suited for
separating the computation}. Better suited for these are the task graphs commonly used for scheduling on multicore systems, or dataflow graphs from models of computation like Synchronous Data Flow (SDF)~\cite{leesdf} 
or Kahn Process Networks (KPN)~\cite{kahn74}. 

The problem of calculating the group of automorphisms of a graph is an extensively studied subject. %
\revi{%
    In practice, it is solvable efficiently for the majority of graph-classes and there are many specialized algorithms/frameworks like bliss \cite{junttila2011}, conauto \cite{lopez2011conauto}, saucy \cite{codenotti2013saucy}, nauty and Traces \cite{nauty} to name a few. %
For example, for a random graph of up to $10^3$ nodes, they are able to determine its automorphism group in less than $0.01$s \cite{nauty}.
All of these algorithms use modern variations of the Schreier-Sims algorithm to manage the graphs' automorphism groups.
}
Computing the automorphism group of a graph is \revi{directly} related to the Graph-Isomorphism-Problem,
\revi{%
a problem in complexity theory famous for being apparently easier than NP-complete problems, %
while its complexity status is not yet settled.
It has recently been claimed to be solvable in quasipolynomial time~\cite{babai2015graph} and is currently being peer-reviewed.}

The following fact is a reinterpretation of the definition of a graph's inverse semigroup of symmetries,
which we use to calculate it.
\begin{fact}\label{fact:partial_symmetry}
  A partial permutation of the set of nodes of a graph is in its inverse semigroup of symmetries if and only if it induces an isomorphism of labeled graphs from the subgraph induced by its domain to the subgraph induced by its image.
\end{fact}
Based on this, we designed Algorithm~\ref{algo:inv_semigroups} to find a generating set $S$ of the semigroup of all such partial permutations.
For two graphs $\Gamma = (V,E), \Gamma' = (V',E')$ and a mapping $\varphi: W \rightarrow W'$ from a subset of vertices $W \subseteq V$, to a subset of vertices $W' \subseteq V'$,
\revi{let $\Gamma_{\mid W}$ be the subgraph of $\Gamma$ induced by the vertex set $W$, i.e. $\Gamma_{\mid W} = (W,\{ (v,w) \in E \mid v, w \in W)$. Then,}
we denote by $\varphi_\Gamma: \Gamma_{\mid W} \rightarrow \Gamma'_{\mid W'}$ the induced mapping on the induced subgraphs, i.e. $\varphi_\Gamma( (v_1,v_2) ) = (\varphi(v_1),\varphi(v_2) )$, \revi{as is the natural way of extending the mapping to the edges}.
Note that this is only \revi{a partial automorphism of $\Gamma$}, if $\varphi(e) \in E'$ for all $e \in E$. 

\begin{algorithm}
  \caption{Inverse semigroup of symmetries of a graph}
  \label{algo:inv_semigroups}
  \begin{algorithmic}[1]
    \Input{A graph $\Gamma = (V,E)$}
    \Output{ A generating set of partial permutations $S$ }
    \State $S \gets \emptyset$
    \State $A \gets \langle S \rangle$
    \revi{~~// compute a data structure}
    \State $C \gets \{ \varphi \mid \varphi \text{ is a partial permutation of } V \}$
    \ForAll{ $\varphi \in C$ }
    \If {$\varphi_\Gamma$ is a partial automorphism}
    \If {$\varphi \notin A$}
    \revi{~~// membership test}
    \State $S \gets S \cup \{ \varphi \}$
    \State $A \gets \langle S \rangle$
    \revi{~~// update the data structure}
    \EndIf
    \EndIf
    \EndFor
    \Return $S$
  \end{algorithmic}
\end{algorithm}

The correctness of Algorithm~\ref{algo:inv_semigroups} follows from Fact~\ref{fact:partial_symmetry} and its termination from the fact that the set of all partial permutations of a set is finite.
Our implementation uses the computer algebra system GAP~\cite{GAP4} and specifically its semigroups-package~\cite{semigroups} \cite{east2015}
\revi{%
for performing constructive recognition and membership testing in Algorithm~\ref{algo:inv_semigroups}, lines 2, 6, 8.
}

It is easy to see that Algorithm~\ref{algo:inv_semigroups} is not optimal, since it considers redundant generators. For example, consider two sets $V_1 \subsetneq V_2 \subseteq V$. For all partial functions
\revi{$f \in \langle S \rangle$} which have $V_2$ as its domain, the partial function $f_{\mid V_1}$ is also in $\langle S \rangle$.
\revi{%
As a first improvement, we developed a backtrack-based algorithm sketched in Algorithm~\ref{algo:backtrack}.
We utilize that, if a partial permutation is not an isomorphism, neither can its extensions. %

The recursive backtrack loops over the set of all partial permutations by organizing the partial permutations into a tree, and traversing it depth-first.
The tree, in which each node represents one partial permutation of $V$, is built up as follows:
Let the root of the tree be the empty partial permutation $\varepsilon \colon \{\} \to \{\}$.
The children of a node $\varphi$ consist of all partial permutations $\varphi^\prime$ that extend $\varphi$ by exactly one further point.
E.g.\ one child of the node $\pi \colon 1 \mapsto 4, 3 \mapsto 2$ would be the node $\pi^\prime \colon 1 \mapsto 4, 3 \mapsto 2, 4 \mapsto 1$.

We can improve this algorithm, by initializing $S$ with partial automorphisms we can readily deduce from the NoC architecture, like the global symmetries from the group of automorphisms.
There are two further technical improvements we need to make Algorithm~\ref{algo:backtrack} feasible.
On the one hand, we need to make sure we do not perform the membership test for all restrictions of a partial automorphism $\varphi$, but only for $\varphi$ itself.
On the other hand, the tree contains most partial permutations several times,
e.g.\ a partial permutation $\varphi$ with domain $\{ 1,3,4 \}$ would appear as a child of both a node $\varphi_1$ with domain $\{ 1,3 \}$ and a node $\varphi_2$ with domain $\{ 1,4 \}$.
Thus, we need to make sure that each partial permutation appears exactly once. %
}

\revi{
\begin{algorithm}
  \revi{
  \caption{Inverse semigroup of symmetries of a graph with backtracking}
  \label{algo:backtrack}
  \begin{algorithmic}[1]
    \Input{A graph $\Gamma = (V,E)$}
    \Output{ A generating set of partial permutations $S$ }
    \State $S \gets \emptyset$
    \State $A \gets \langle S \rangle$
    \State $\varepsilon \gets$ empty partial permutation
    \State $\text{backtrack}(\varepsilon)$
    \State \Return $S$
  \end{algorithmic}
  \begin{algorithmic}[1]
    \Functionname{backtrack}
    \Input{A partial permutation $x$}
    \If{$x_G$ is not a partial automorphism}
      \State \Return
    \EndIf
    \If { $x \notin A$ }
      \State $S \gets S \cup \{ x \}$
      \State $A \gets \langle S \rangle$
    \EndIf
    \ForAll{ $c \in \text{children}(x)$ }
      \State $\text{backtrack}(c)$
    \EndFor
  \end{algorithmic}
}
\end{algorithm}

  }
\subsection{Domain-specific optimizations}

The standard methods from algorithmic group theory are designed to be general and include all corner cases. In the case of applications for software synthesis, this is not necessary.
While we used existing standard algorithms or simple naive algorithms to evaluate the applications, we can easily identify potential for domain-specific optimizations in these areas.
In the future we will investigate specially-tailored algorithms that leverage assumptions about the possible symmetry groups. Using this domain-specific knowledge is bound to yield much better time-efficiency results.

For example, by using structural properties of groups, a dynamic-programming approach can be used to substantially reduce the memory footprint of the methods \cite{lubeck2001enumerating}.

\revi{Another} possible domain-specific optimization would rely on the fact that hardware architectures, while varied, mostly have a limited amount of complexity. 
It would thus be possible to create a classification of common groups in the architectures and applications. This works particularly well since the nature of the symmetry groups for architecture graphs is compositional.
For example, the group of a automorphisms of a heterogeneous bus-based architecture with $k$ different PE types is the direct product of the collection of symmetric groups $S_{n_i}$, where the architecture has $n_i$ PEs of type $i$, 
which is the group of automorphisms of a homogeneous bus-based architecture for the respective PE types.

Finally, Algorithm~\ref{algo:backtrack}still has to iterate over the whole semigroup of automorphisms, which, even for the 4x4 NoC already contains around 1.2 million partial symmetries.
Since the NoC architecture graphs have a very regular structure, we are confident that, in future work, we can leverage it and develop a specialized version for such NoC architectures.
Even more so, it might be possible to classify the semigroups of automorphisms of such architectures by purely theoretical means, circumventing the need for computations at all.
\revi{%
\subsection{Computational Overhead}
}
\label{sec:complex}

\begin{figure}
	\centering
	\includegraphics[width=0.6\textwidth]{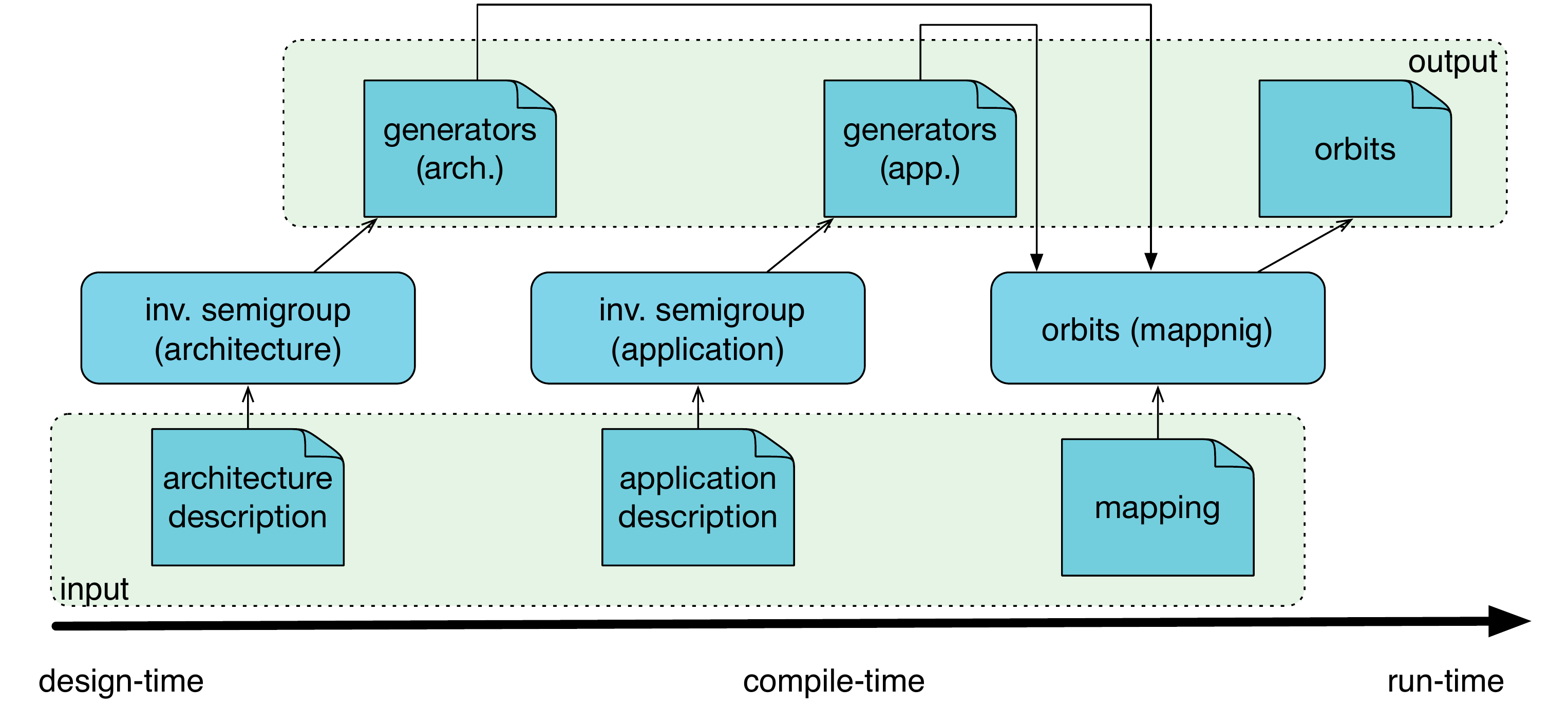}
	\caption{An overview of the different symmetries in software synthesis.}
	\label{fig:symmetries_flow}
\end{figure}

As we have seen in this section, using the methods of group and inverse semigroup theory at diverse stages of the software synthesis process, will inevitably lead to computational overheads.
This leads to a trade-off: while knowing that two mappings are equivalent can have benefits in terms of computation time, this requires a computational effort.
It is not evident that these methods will be computationally beneficial in every use case. However, there are several reasons why such use cases are bound to be common.
Some parts of the calculation, like the symmetry group of the architecture, can be executed only once, at design time for a hardware architecture for instance.
They can subsequently be and used on every compilation that targets the architecture.
Figure~\ref{fig:symmetries_flow} illustrates this by giving an overview of the different components of symmetry in software synthesis and their time-point at which they can be leveraged.
In many use-cases, a complex and time-costly simulation is executed for every mapping. By leveraging the fact that symmetry calculations only depend on the structure of the problem,
the complexity of the simulation can be increased with minimal penalty, since it has no effect on the symmetry calculations.

\section{Evaluation}
\label{sec:eval}
In this section we present two use cases using the proposed symmetry principles for reducing design space exploration.%
We use the principles to enhance existing tools and evaluate the usefulness of considering symmetry in software-synthesis, or more specifically, in design-space exploration. %
\subsection{Use Case: Genetic Algorithms}
\label{sec:improving_evo}
The first use case deals with the scalability of a metaheuristic based on genetic-algorithms which maps tasks to processing elements.
Concretely, we consider a design-space exploration suite of the Sesame framework~\cite{sesame}, and use symmetries to prune the search-space.

Sesame is an open-source framework that tackles the problem of efficiently executing parallel applications on heterogeneous MPSoCs. 
By structuring programs as Kahn Process Networks (KPNs)~\cite{kahn74}, Sesame gathers execution information in the form of process traces.

Using high-level hardware models, Sesame can use execution traces to drive an accurate albeit efficient simulation of the execution of the KPN application.
The high-speed, high-fidelity of this trace-based simulations permits to explore the design space of \revi{static (run-time)} mappings by using so-called metaheuristics, like evolutionary algorithms \cite{erbas_pimentel06}.
Metaheuristics are a general approach to solving optimization problems with large design spaces which exhibit structure. 
They are based on the principle of iterating over different solutions in order to improve them and reach near-optimal results. 
In the case of evolutionary algorithms, the method for iteration is inspired in the biological process of evolution.
Solutions are called individuals, and they are recombined to produce offspring, also introducing random mutations, over several generations.

Due to the iterative nature of metaheuristics, they rely on several simulations of the application using different mappings. 
These simulations are computationally expensive, especially so for larger applications. 
In most cases the simulation time dominates the total exploration time.

\begin{figure*}[t!]
	\subfloat[Improved Sesame flow for genetic algorithms.\label{fig:sesame_symmetries}]{
     \begin{minipage}[b]{0.48\textwidth}
	\includegraphics[width=1\textwidth]{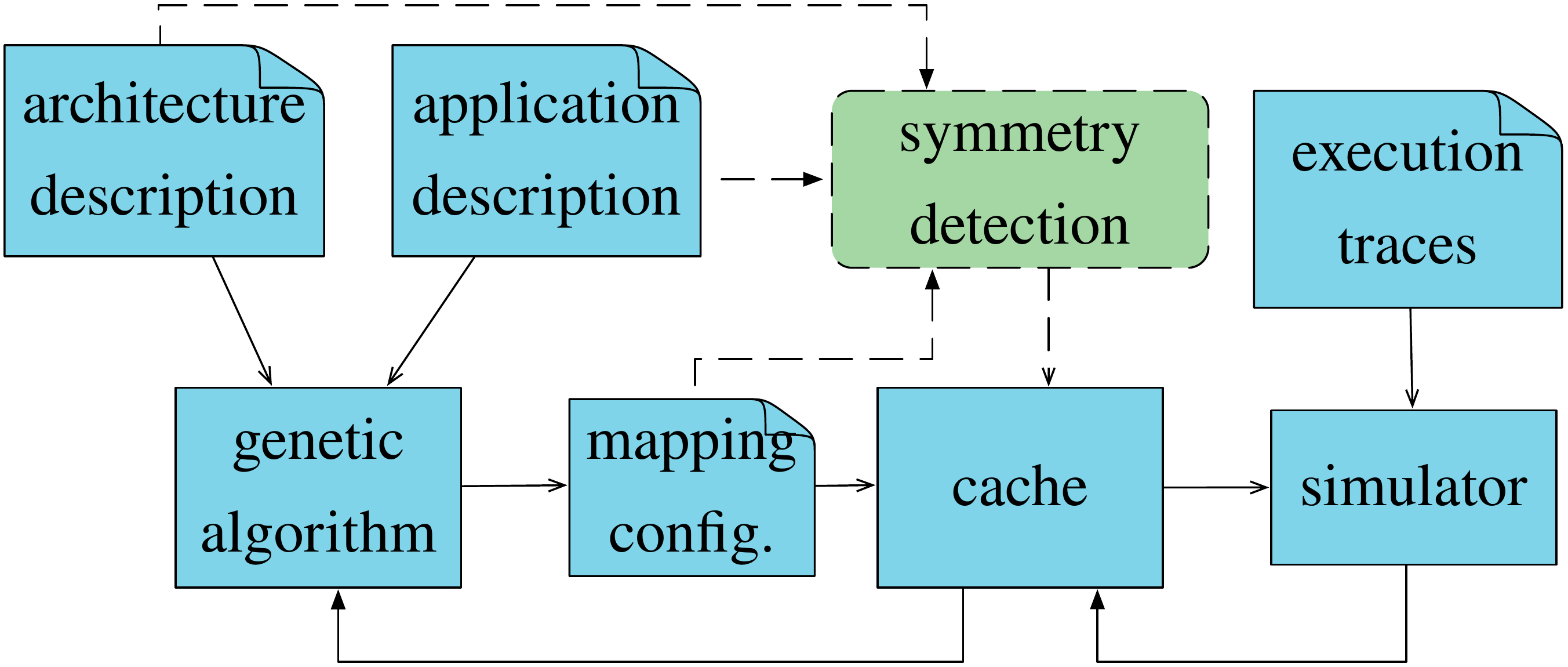}
 \end{minipage}
   }
	\hfill
	\subfloat[Improved sub-architecture exploration.\label{fig:platform_grower_symmetries}]{
\begin{minipage}[b]{0.48\textwidth}
	\includegraphics[width=1\textwidth]{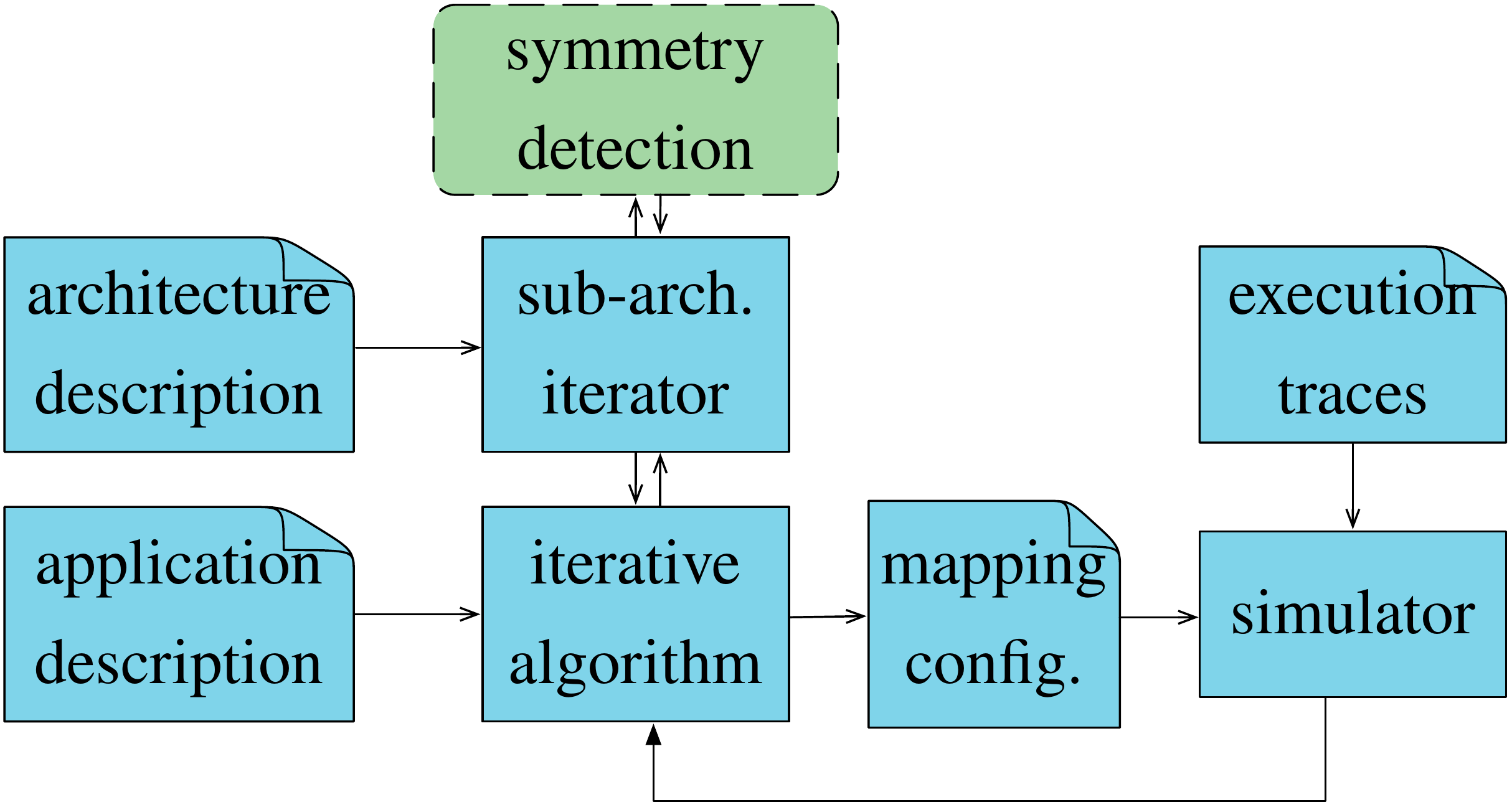}
\end{minipage}
}
\caption{A high-level overview of the use cases with our symmetry improvements.}
\end{figure*}

The mutation and cross-over properties of genetic algorithms (GA), a type of evolutionary algorithms, are likely to produce similar mappings in large populations.
The DSE framework includes a cache which substitutes a simulation with a look-up when a mapping has already been simulated.
To improve the execution time and scalability of the evolutionary algorithm, we propose extending this caching strategy.
Figure~\ref{fig:sesame_symmetries} shows a high-level overview of the Sesame flow using genetic algorithms for mapping and our proposed solution.
In our extended caching strategy, the simulation is only executed once for all \revi{equivalent} mappings (see Section~\ref{sec:actionsonmappings}).
This significantly reduces the search space, since it replaces the space of valid mappings with the space of equivalence classes of valid mappings.
\revi{The python implementation of this GA uses a hash-map as a cache.
  Since we only need ensure we do not simulate a mapping if we have simulated an equivalent one before, we implemented this by using a canonical representative (see Section~\ref{sec:actionsonmappings}) of the class as the key.
This is significantly more efficient than having to calculate the whole orbit when orbits become large, i.e. in the case when there is much symmetry.}

\subsection{Use Case: Sub-Architectures}
\label{sec:improving_fun}
The second use case considers an iterative heuristic for minimizing the usage of system resources in in the presence of application constraints.
For this, we use an iterative heuristic from a modern, commercial multicore compiler, from the SLX Tool Suite~\cite{slx}. Similar algorithms have been presented in~\cite{Cheung07,castrillontii13}.

Instead of finding the system with maximal performance, in several scenarios application developers are interested in using the least resources possible
in a system, while meeting some performance constraints. This is the case in hard or soft real-time applications, e.g., in the multimedia and automotive domains.
The iterative algorithm explores mapping configurations using limited resources. \revi{However, these resources might be heterogeneous, which makes the simple approach of the algorithm very limited.
The basic algorithm works as follows:}

\begin{enumerate}
\item Start by considering a sub-architecture consisting of a single PE.
\item \label{step:mapping} With heuristics, calculate a mapping for close-to-optimal performance in the current sub-architecture.
\item Use a fast trace-driven simulation to estimate the performance of the mapping.
\item If the mapping is fast enough in the simulation, finish and return the mapping.
\item \label{step:increase_subarch} If the mapping is too slow, consider an additional, \revi{randomly-chosen} PE in the sub-architecture, as well as the corresponding communication resources.
\item Return to Step~\ref{step:mapping}.
\end{enumerate}

This algorithm can be very ineffective on heterogeneous platforms.
\revi{
The concrete sub-architecture considered depends on the choices of PEs taken in Step~\ref{step:increase_subarch}, which is too simplistic.
Thus, the algorithm does not consider several configurations using different PE types that could perform better, and completely disregards the architecture topology. 
For comparison, we will refer to this variant of the algorithm as \textbf{simple}.

On the other hand, considering every possible sub-architecture in a \textbf{brute-force} approach scales prohibitively poorly:} an architecture with $n$ PEs
has $2^n - 1$ possible (non-empty) sub-architectures, even disregarding communication. This is considerably larger than the $n$ sub-architectures considered by 
iteratively adding PEs.

To deal with this issue, we propose to use the architecture symmetries to select the relevant sub-architectures to be considered in the iterative algorithm.
We choose a single representative of every equivalence class of the architecture, before mapping.
We then modify the iterative algorithm. Figure~\ref{fig:platform_grower_symmetries} sketches the flow using the improved algorithm. 
It is clear from the figure that the symmetry reduction step can be precomputed, since it only depends on properties of the target architecture.
Instead of just increasing the size of the sub-architecture, our modified algorithm iterates over all non-equivalent sub-architectures of a given size first.
If none of the non-equivalent sub-architectures of size $n$ give enough performance, then sub-architectures of size $n+1$ are considered.
This modification improves the algorithm by letting it acknowledge the heterogeneity of the architecture in a reasonable time. 
The exact number of equivalence classes depends on the architecture and the inverse semigroup describing its symmetry.
\revi{In this context, we differentiate two additional strategies: \textbf{groups}, which uses the symmetry group to differentiate subarchitectures, and \textbf{inv. semi.}, which uses the inverse semigroup instead.}

\subsection{Results}
\label{sec:evaluation}

\begin{wrapfigure}{r}{0.6\textwidth}
	\centering
	\includegraphics[width=0.6\textwidth]{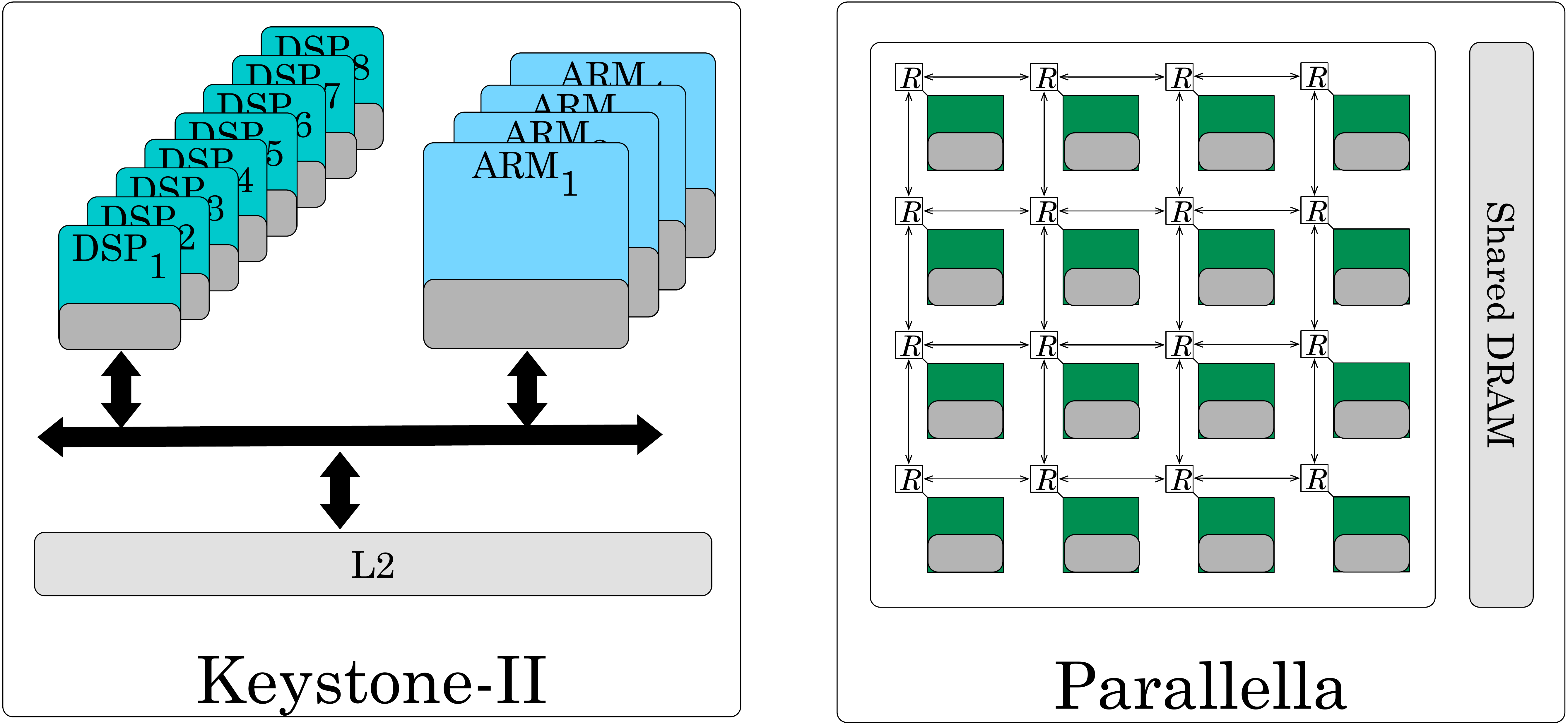}
	\caption{Schematics of the Architecture Models}
	\label{fig:architectures}
   \end{wrapfigure}
We test both example use cases with several benchmarks from the signal processing and multimedia domains.
For this, we use current applications and real architectures.
While we can only expect little gains in small applications mapped into architectures with a handful of cores,
these serve to evaluate the symmetry methods with realistic parameters.

\subsection{Genetic Algorithms}

Table~\ref{tab:benchmarks_evo} gives a summary of the benchmarks used, and in particular, 
the number of processes of each benchmark application, the isomorphy type of the symmetry group and its size. 
\revi{The symmetry groups of the applications were determined by hand, by inspecting the code of the benchmark.
Automating this task in a compiler is highly non-trivial in general and is outside the scope of this paper.}
We tested the applications by mapping on an accurate model of a state-of-the art heterogeneous multicore system, the Texas Instruments (TI) Keystone II. It features 4 ARM Cortex-A15 processing elements and 8 DSPs~\cite{keystone2_whitepaper}. 
The model was adapted~\cite{goens_mcsoc16} from a model of the commercial state-of-art MPSoC compiler from the SLX Tool Suite and confirmed with measurements from hardware~\cite{odendahl2013split}.
A schematic view of the architecture can be seen in Figure~\ref{fig:architectures}.

\begin{figure}
	\centering
	\includegraphics[width=1\textwidth]{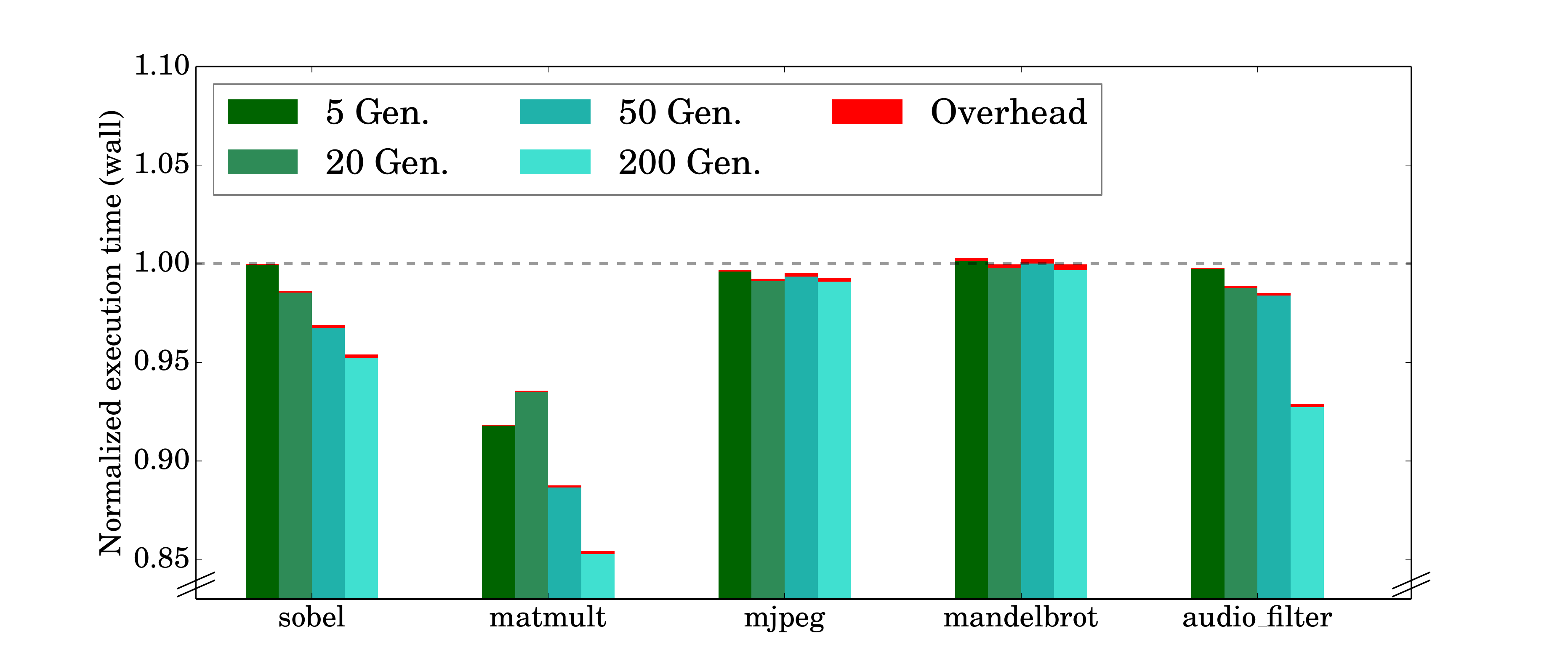}
	\caption{A comparison of benchmarks for the improved caching strategy in Sesame. The red bars on top represent the overhead, while the dotted gray line marks the baseline (without symmetries). }
	\label{fig:genetic_algorithms}
\end{figure}

The TI Keystone II has a symmetry group of size $4! \cdot 8! = 967680$. It is isomorphic to the direct product of the symmetric groups on $4$ and $8$ points, usually denoted by $S_4 \times S_8$.
For this example, the symmetry group already yields all symmetries, as motivated in Section~\ref{sec:groupsnotenough}.
This is the case since the simple interconnect topology (bus-based) does not break the symmetry, only the different core types do. 

\begin{table}
\caption{Benchmarks and symmetry properties. $C_n$ denotes the cyclic group of order $n$.}\label{tab:benchmarks_evo}
{\scriptsize
\begin{tabular}{lllll}
\hline Name & Short description & No. of pr./ch. & Group & Group Size \\
\hline \hline 
\revi{sobel}&\revi{Sobel filter on $40\times40$ image}&\revi{$5/15$}&\revi{$C_2$}&\revi{2}\\
\revi{matmult}&\revi{$10\times10$ matrix multiplication}&\revi{$5/6$}&\revi{$\{1\}$}&\revi{$1$}\\
\revi{mjpeg}&\revi{Motion JPEG decoder ($128\times128$)}&\revi{$12/15$}&\revi{$S_4$}&\revi{$4! = 24$}\\
\revi{mandelbrot}& \revi{Mandelbrot set calculation ($16$ Jobs)}&\revi{$18/32$}& \revi{$\{1 \}$}& \revi{1} \\
\revi{audio filter}&\revi{Stereo filter from Figure~\ref{fig:audio_filter}}&\revi{$8/8$}&\revi{$C_2$}&\revi{2}\\
\hline
\end{tabular}
  }

\end{table}

\revi{
We executed the DSE framework with $5, 20, 50$ and $200$ generations for all benchmarks, with five different (fixed) random seeds, repeating each such configuration five times.
It uses an evolution-strategy called $\mu + \lambda$, which we used with fixed populations of size $20$ and with $20$ children.
We measured the overall wall-clock time of the DSE using our improved cache strategy and the conventional cache. Additionally, we
measured explicitly the time spent in symmetry-related calculations in the improved cache. The results are summarized in Figure~\ref{fig:genetic_algorithms}.
It shows the normalized average of the total execution time for all benchmarks, separating the time spent on the DSE from the overhead of the symmetry calculations. 
The plot shows the overhead is negligible in all cases; indeed, it represents between $0.1-0.2 \%$ of the total execution time across the different benchmarks.
The plot also shows that the improved cache results in a net improvement of the execution time, in spite being a prototype implementation. }
As expected, the effects of symmetry were more visible in explorations with a larger number of generations (and thus, of simulations).

\begin{figure}[b]
  \centering
  \includegraphics[width=0.62\textwidth]{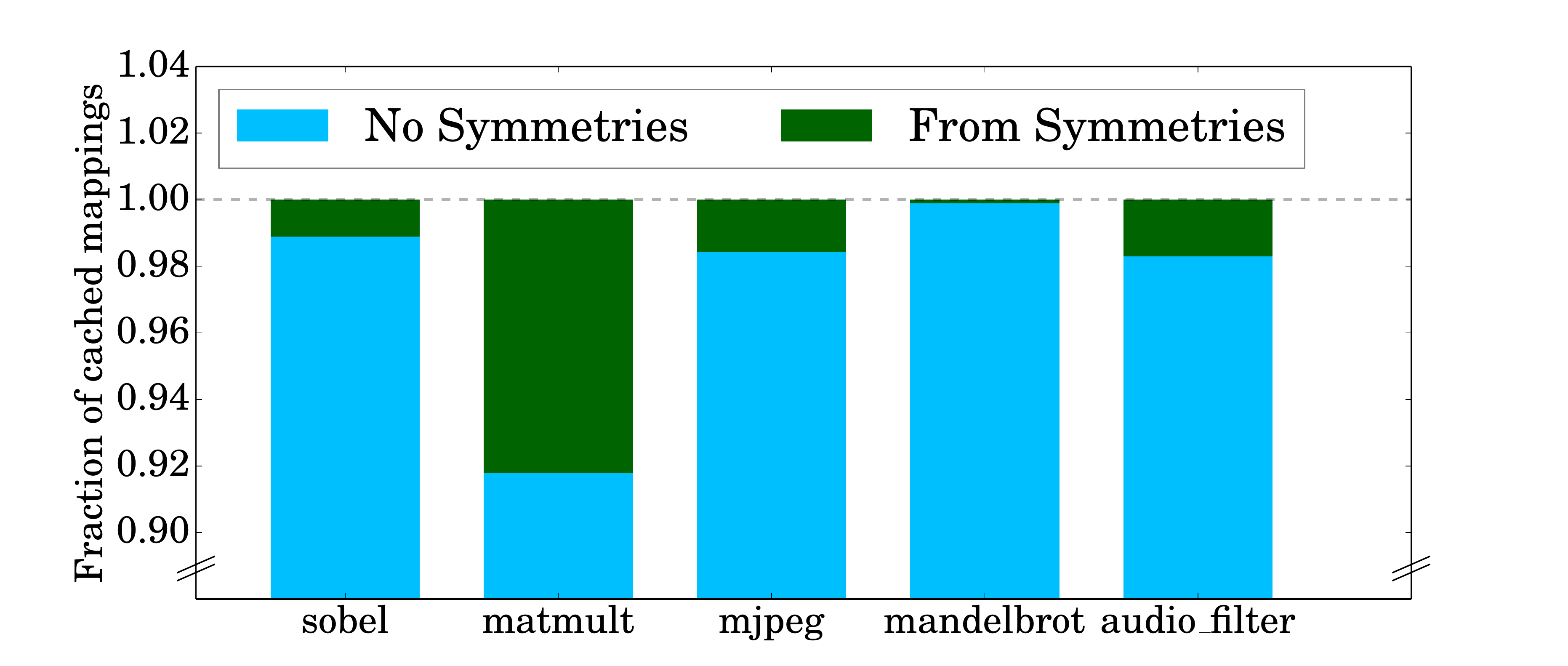}
  \caption{Percentage of cached mappings which were found using symmetries.}
  \label{fig:sesame_cached_percentages}
  	\vspace{-3mm}
 \end{figure}
 \revi{
Additionally, we compared the number of cache hits in the original, unmodified cache and our symmetry-improved version, as can be seen in Figure~\ref{fig:sesame_cached_percentages}.
We see that a large majority of hits happen because exactly the same mapping is being calculated, instead of an equivalent one.
This hints to the fact that the variation in the space is small, and exacerbates the problem that the space itself is not being reduced directly.
Indeed, removing the symmetries in the cache still leaves them in the chromosome models, which technically speaking is not a reduction of the search space, but rather just a run-time optimization of the DSE.
Thus, an in-depth improvement improvement would include symmetry in the chromosome models, letting them characterize only the representatives of the orbits, instead of all mappings.
In particular, since our strategy just improved the caching, the Pareto fronts of best mappings resulting from exploration stay unchanged. By factoring out the symmetry at the chromosome level, we should also be able to produce better mappings in less time.
}
\subsection{Sub-Architectures}
For the second use case, we used a different architecture. The Parallella\textregistered~architecture commercialized by Adapteva includes $16$ Epiphany~\cite{epiphany} cores,
each with its scratchpad memory, and a shared DRAM. We use this architecture to showcase the effects of considering communication in NoC-based architectures and inverse semigroups.
The architecture model is part of the commercial SLX Toolsuite\textregistered~MPSoC compiler and is designed for the high-speed, high-accuracy requirements of this use case.
As application we use an mjpeg decoder, described in~\cite{sheng2012fifo}. 

The four by four mesh structure in the Parallella architecture has $2^{16} - 1 = 65535$ possible sub-architectures. 
In contrast, the architecture has only $8547$ equivalence classes of (non-empty) sub-architectures considering the symmetry group, and these are in turn reduced to $6803$ if we consider the inverse semigroup. 
In most use cases, not all sub-architecture sizes would be used. Instead, only architectures up to a particular size would be considered. 
\begin{figure}
    \centering
    \includegraphics[width=0.7\textwidth]{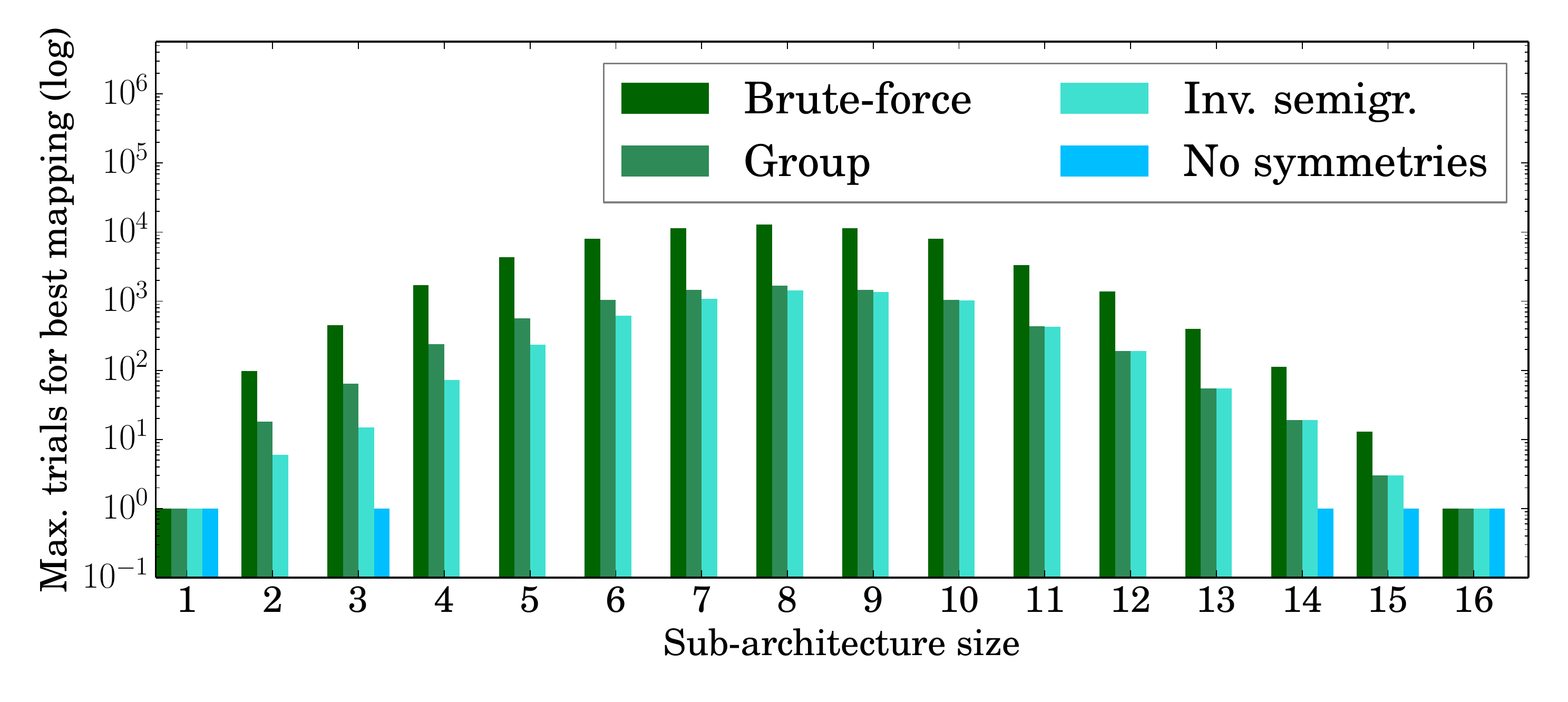}
    \caption{Maximal number of trails for obtaining the best mapping using the different strategies. The missing bars on the ``simple'' strategy represent failures of the strategy to find the best mapping.}
    \label{fig:explorer_trailstomax}
\end{figure}

\revi{We evaluated the four variants of the strategy for choosing sub-architectures described in Section~\ref{sec:improving_fun}: simple, groups, inverse semigroups (inv. semi.) and brute-force. }
The results obtained with all four variants of the algorithm can be seen in Figure~\ref{fig:explorer_trailstomax}. 
The plot depicts, for every sub-architecture size, the maximal number of trails required to obtained the best mapping for that sub-architecture size. Several bars for the ``simple'' strategy cannot be seen in the plot since,
in most cases, this strategy did not find the best mapping for a given sub-architecture size.
\revi{
In those cases, the best mapping found with the ``simple'' strategy was off by an average margin of $2\%$. However, for smaller sub-architecture sizes, which are usually the ones we are interested in,
the error was mostly around $7-10\%$ in the experiments.}

\revi{ Note that the benefit of using inverse semigroups over groups decreases for larger sub-architecture sizes. This is a consequence of the nature of the inverse semigroup of symmetries, which is composed of \emph{partial} permutations. When
the sub-architecture becomes large enough, many partial symmetries do not apply anymore, only \emph{global} symmetries do; those already captured in the concept of the symmetry group.}

\begin{figure*}[t!]
	\centering

	\subfloat[Comparison for a fixed size ($4$).\label{fig:fixedsize}]{
		\begin{minipage}[b]{0.48\textwidth}
	\centering
	\includegraphics[width=1\textwidth]{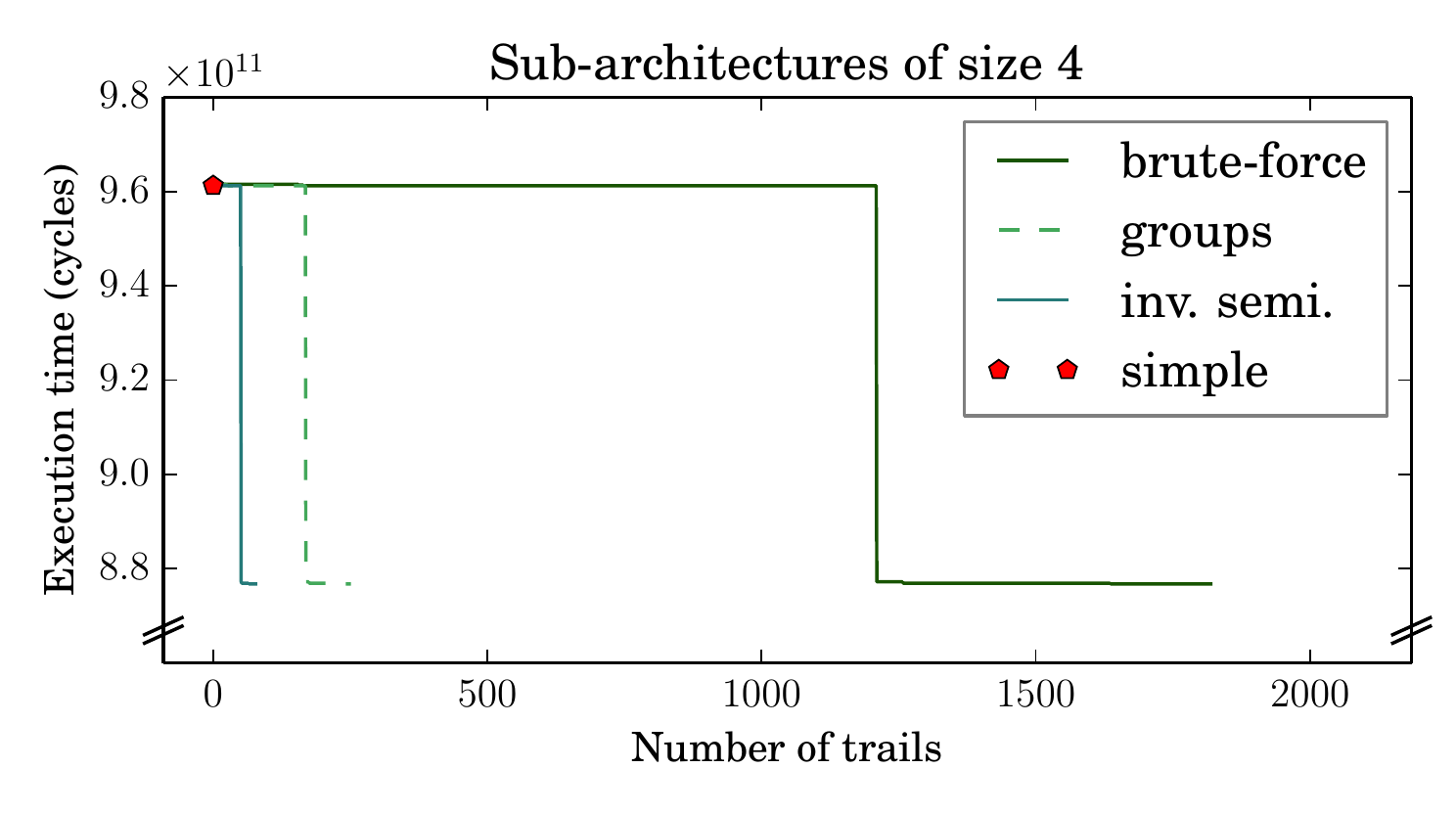}
		\end{minipage}
    }
	\subfloat[Quality of the solutions.\label{fig:realtime_algorithms}]{
	\hfill
		\begin{minipage}[b]{0.48\textwidth}
	\centering
	\includegraphics[width=1\textwidth]{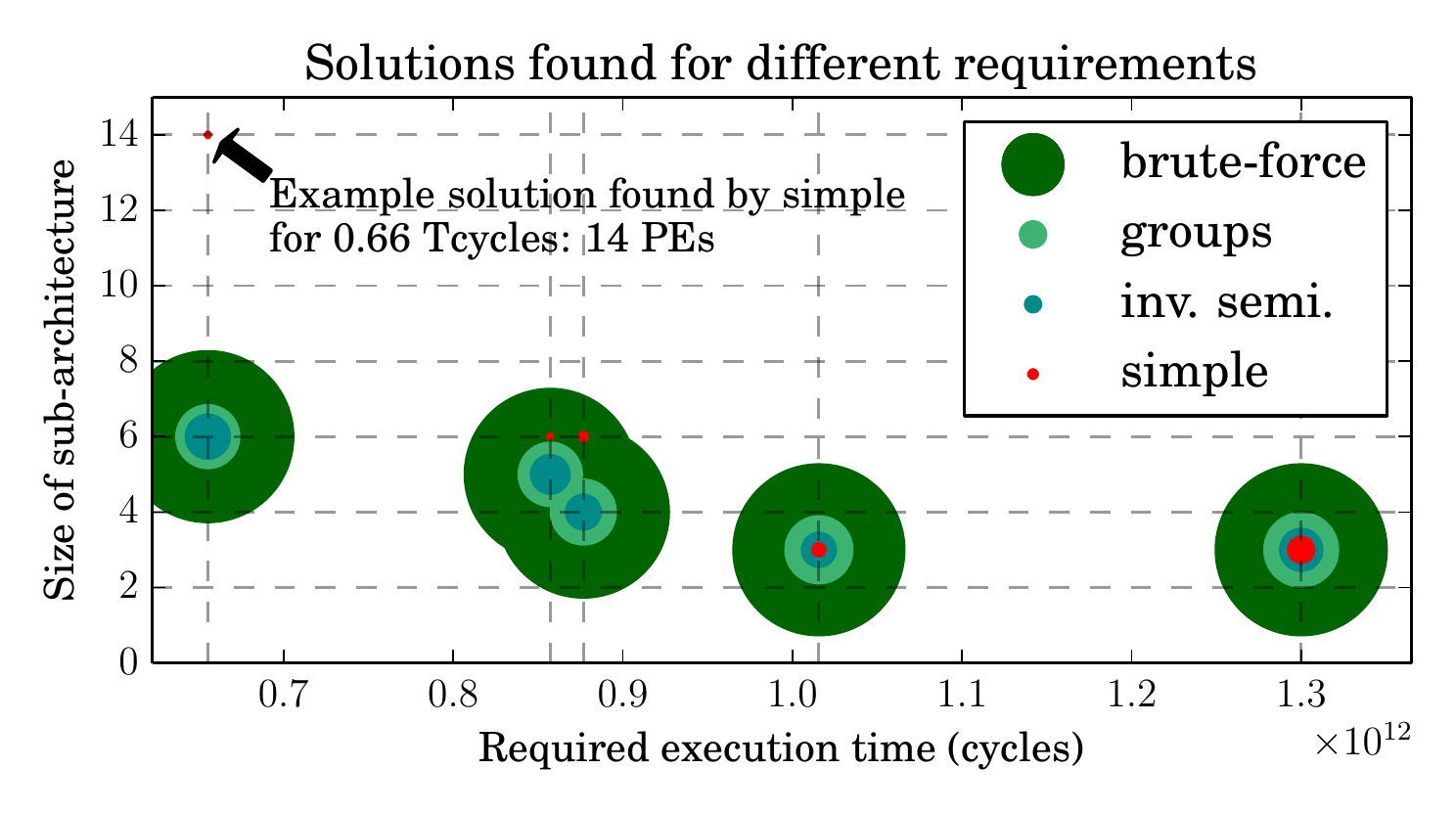}
		\end{minipage}
	}
	\caption{Comparison of strategies for resource-constrained mapping.}
	\label{fig:realtime_full}
	\vspace{-3mm}
\end{figure*}

A different perspective of the results can be obtained from Figure~\ref{fig:fixedsize}. 
\revi{
It shows a series of sub-architecture configurations,  where every point represents an evaluation for a different sub-architecture.
All sub-architectures have $4$ PEs, and the evaluations are sorted in descending order by the simulated execution time. 
Consider the first trial in Figure~\ref{fig:fixedsize}. All three strategies find similar mappings of $9.6 \cdot 10^{11}$ cycles. The inverse semigroups strategy finds the better mappings with $8.8 \cdot 10^{11}$ cycles after $50$ trials,
whereas the group-based strategy needs over $160$. The simple strategy never finds these better mappings. Assuming this application had a requirement of running under $9\cdot 10^{11}$ cycles, the simple strategy would need more than $4$ PEs to achieve this. 
In fact, the simple strategy only found an application running under $9\cdot 10^{11}$ cycles after considering $6$ PEs, as can be seen in Figure~\ref{fig:realtime_algorithms}.
It shows, for different run-time requirements, the number of PEs in the best sub-architecture found with the different strategies.
The relative sizes of the dots represent the relative number of trials to find that sub-architecture. Even worse than the example with $9 \cdot 10^{11}$ cycles is the example marked in the figure, with $6.5 \cdot 10^{11}$ cycles. Instead of the $6$ PEs required with the
mappings found with the ``groups'', ``inv. semi'' and ``brute-force'' strategies, the ``simple'' strategy requires $14$ PEs. This shows how the quality of the solutions can be improved significantly with our proposed methods.
}

Even for the modestly-sized Parallella architecture, the brute-force algorithm requires at least ten times as many iterations as the other algorithms.

\begin{figure}
	\centering
	\includegraphics[width=0.9\textwidth]{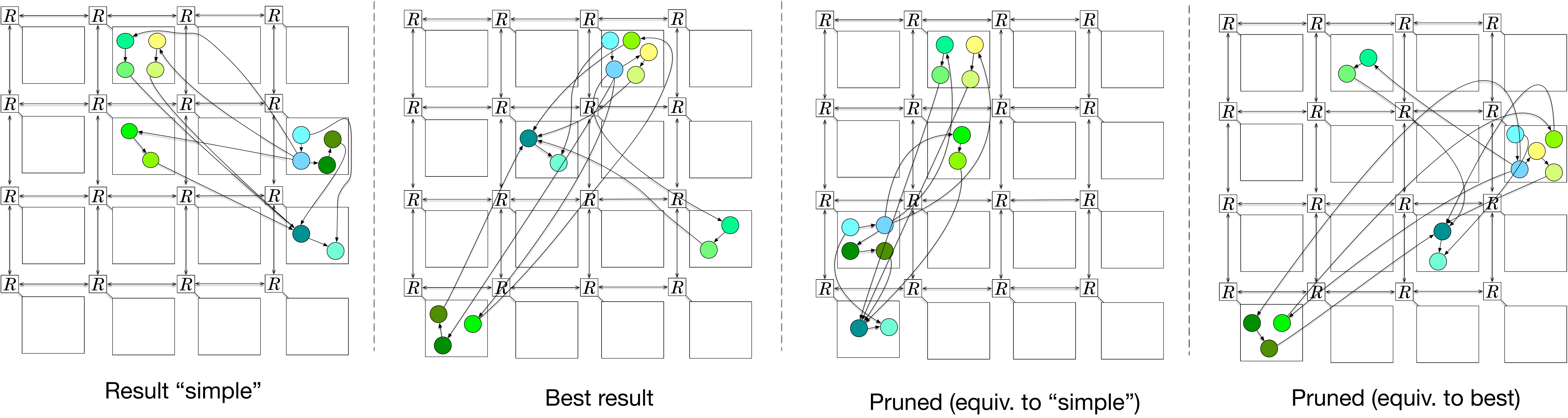}
	\caption{Examples of mappings of the mjpeg application to sub-architectures of size $4$ of the Parallella, colored to help identify the processes. Symmetries become less evident with growing complexity. }
	\label{fig:mappings_subarchs}
\end{figure}

Finally, Figure~\ref{fig:mappings_subarchs} depicts concrete mappings from Figure~\ref{fig:fixedsize}. The processes and the FIFO buffers of the mjpeg application
are depicted as a graph, embedded on the architecture. The location of the processes in the figure represents the PE to which they are mapped.
The mapping labeled with ``Result simple'' in Figure~\ref{fig:mappings_subarchs} is precisely the single red dot from the ``simple'' strategy in Figure~\ref{fig:fixedsize}. The mapping labeled as ``Best result'' is one of the mappings with
the lowest execution time found by the ``inv. semi.'' strategy in Figure~\ref{fig:fixedsize}. Notice that the communication seems to be sub-optimal, but from the critical path for this execution, in this particular benchmark this had no effect: 
moving the ``Best result'' mapping to contiguous PEs yielded the same execution time in the simulation. The other two mappings show two sub-architectures which were pruned by our method, the one labeled with ``Pruned (equiv. to simple)'' is equivalent to the 
first mapping, ``simple'', i.e., they both produced the same results, which was identified by our method. The situation is similar for the second and the last depicted mappings. 
\revi{Note how, while the communication distances between processes are always preserved, the transformations cannot be given simple geometrical interpretations, as ``shapes''.}
This figure serves to show concretely how the limits of the intuitive symmetry concepts can be quickly reached, and why an automated approach like ours is necessary.

\revi{Finding the complete (generating set of the) inverse semigroup took about 30 min on an Intel Xeon E5-4617 running at  $\unit[2.90]{GHz}$, with Algorithm~\ref{algo:backtrack}, and using the $12$ generators obtained, it took slightly over $3$ seconds to calculate all representative sets.}
This time is negligible compared to the synthesis time of the $58732$ mappings for sub-architectures that were pruned, 
which would represent over $34$ days of computation on the system used to evaluate this use-case.
Additionally, this computation was done once, and can be used with basically no overhead for any software synthesis to the same architecture. 

\revi{
As a final summary, Table~\ref{tab:summary_evaluation} gives an overview of the gains from symmetry and overheads in the different benchmarks for the two use-cases. We see that, even for the small architectures and applications of today,
prototype implementations already yield significant improvements. 
}

\begin{table}
  \revi{
\caption{Summary of gains and overheads from symmetry in our use-cases. All relative numbers are normalized with the (average) baseline benchmarks without symmetries.}\label{tab:summary_evaluation}
{\scriptsize
\begin{tabular}{lllll}
\multirow{2}{*}{Use-case}& \multirow{2}{*}{Benchmark} & \multicolumn{2}{l}{Relative overhead (avg.)}  & \multirow{2}{*}{Relative time-reduction} \\
  \cline{3-4} 
 & & Design-time& Run-time & \\
\hline 
Genetic algorithms & sobel        & $\approx 0$ &  $(0.11 \pm 0.04)\%$& $(2.38\pm 2.05)\%$\\ 
\phantom{Genetic algorithms} & matmult      & $\approx 0$ & $(0.095\pm 0.04)\%$& $(10.2\pm 3.6)\%$\\ 
\phantom{Genetic algorithms} & mjpeg        & $\approx 0$ &  $(0.14 \pm 0.04)\%$& $(0.71 \pm 0.24)\%$\\ 
\phantom{Genetic algorithms} & mandelbrot   & $\approx 0$ & $(0.21\pm 0.07)\%$& $(0.1\pm 0.2)\%$ \\
\phantom{Genetic algorithms} & audio filter & $\approx 0$ & $(0.12 \pm 0.05)\%$ & $(2.96 \pm 3.73)\%$\\ 
Sub-architectures & groups & $\approx 0$ & $3 \cdot 10^{-7} \%$ & $87.0\%$  \\ 
\phantom{Sub-architectures} & inv. semi &  $0.05\%$ & $3 \cdot 10^{-5}\%$ & $89.6\%$ \\ %
\hline
\end{tabular}
}
}
\end{table}

\section{Related Work}
\label{sec:related}
Besides our two use-cases, a large body of research exists that has addressed hardware and software models for design-space exploration.
\revi{
In particular, the mapping problem and diverse heuristics have been intensively studied before, as can be seen in~\cite{singh2013mapping}.
Besides works that optimize for resources under run-time constraints or multi-objective meta-heuristics, the use-cases studied in this paper,
several other different objectives for mapping have been studied, usually without considering symmetries.
For example, the authors in~\cite{khdr2015thermal} consider thermal-aware application mappings, inspired by dark silicon, but their approach does not leverage the problem's symmetries.
Similarly, mappings aimed at achieving reliability via redundancy have been studied as well~\cite{chen2016task}, \ag{if it comes out on time: robustness with DC}
or motivated by security, as the authors in ~\cite{weichslgartner2016design} present. }

In general, most authors consider the symmetries of the problem in an intuitive fashion~\cite{roloff2015execution,varghese2015programming,thompson2013exploiting,kreutz2005design,hannig2001design,singh2013accelerating}. 
The prevailing theme in all these approaches is that the methods are not general. They are restricted to a certain architecture topology or class of topologies considered by the authors, or particular to the application~\cite{cohen1988symmetry}. 
For example, the authors in \cite{singh2013accelerating} define a model for synthesizing NoC-based MPSoCs on reconfigurable hardware. In their model they define an architecture
composed of tile-types and with a property ``hop\_distance'', which they use to calculate communication costs in a heuristic, by using the value of the largest hop\_distance in a pessimistic fashion. 
While this model seem generic at first, it indeed refers only to a restricted family of architectures, since includes several assumptions about the platform.
It assumes the platform has only one basic communication resource and can be parametrized solely by the hop\_distance and the types of tiles.
This leaves out exotic NoC topologies, with clustering and heterogeneous memories, or whatever the future might hold. Additionaly, while the heuristic might be useful in many use cases,
for those where it is not, the approach has to be redesigned from start to account for the symmetry in a more precise fashion. 
\revi{Similarly, in~\cite{thompson2013exploiting}, the authors explicitly design a method of removing the symmetry as presented in this paper, albeit for a very particular use-case: fully-symmetric homogeneous architectures.
}
\revi{
Finally, the approach in ~\cite{weichslgartner2016design} does recognize and leverage some symmetries explicitly,
namely those which are induced by geometric transformations (rotations, reflections and translations). While this is a more explicit exposition of symmetries, it is limited to 
a very particular architecture family (regular NoC architectures), and will fail to find many symmetries, all those that do not correspond to this intuition; this is specially problematic when the notions of
(euclidean) geometric shapes break down outside of this very particular type of architecture.}

On the other hand, works on high-level design-space exploration~\cite{palermo2005multi,liu2012compositional,quan2014biasedelitist} do not consider symmetries in the design space per-se, 
which implicitly implies that the designer has to come up with clever models for every particular problem. 
Research efforts have also concentrated directly in the reduction of the design-space \cite{wang2004design}.
However, these works achieve a reduction based on properties of the data, not with domain-specific knowledge obtained from application and hardware symmetries.

On the side of group theory, it has a plethora of applications outside of mathematics. It is ubiquitous in the fields of crystallography and particle physics, for example. 
In~\cite{goens_iess15} we considered this issue and primitive version of this framework based on group theory as well.
The theory of inverse semigroups seems to find applications mostly in other fields of mathematics~\cite{lawson1998inversesemigroups}. A similar approach,
based on the abstract concept of groupoids has found other applications, e.g. in theoretical physics \cite{weinstein1996groupoids} or cell biology \cite{stewart2003symmetry}.
However, groupoids do not allow arbitrary multiplications between elements, which makes it complex to generalize most algorithms from group theory in an efficient fashion. 
We believe this makes our approach using inverse semigroups superior. To the best of our knowledge, this work is the first one to address the full symmetries of arbitrary applications 
and hardware architectures in system-level design in a structured and general fashion.

\section{Conclusions and Outlook}
\label{sec:conclusions}
In this paper we have shown how symmetries can be effectively leveraged in software synthesis.
We introduced a rigorous approach using group theory, as well as its generalization in the form of inverse semigroups.
We built a framework for automatically identifying symmetries and analyzed algorithms for doing so, as well as 
potential for domain-specific optimizations.

As a proof of concept, we applied it to two examples of design-space exploration, concentrating our evaluation on current architectures and applications.
Even for the modestly-sized problems of today, we found that the methods provide an advantage concerning scalability by reducing the design space.
\revi{The results yielded time-improvements in virtually all use-cases, from a few percentage points faster up to a full order of magnitude faster in the tested benchmarks. }

Our exploration of applying the methods of group and inverse semigroup theory to software synthesis revealed a very promising field of application. 
\revi{We have outlined diverse domain-specific improvements that could improve these methods. Additionally, }the methods presented are generic, and can thus be applied to a wide variety of problems in neighboring domains.
For example, careful investigation of other programming models (e.g. OpenMP 4.X) can be used for identifying application symmetries in the proposed formalism and improving their execution.
Similarly, these methods can be used on runtime systems for selecting equivalent variants of an execution, to accommodate multiple applications or deal with thermal issues.

\section*{Acknowledgments}
\small
This work was funded in part by the Center for Advancing Electronics Dresden’ (cfaed) and the Graduiertenkolleg Experimentelle und konstruktive Algebra’ (GK EukA).
We thank Silexica for making their SLX Tool Suite available to us.
We thank Andy Pimentel and Simon Polstra for help with Sesame
and Akash Kumar, Alice Niemeyer and Sebastian Posur for interesting discussions which helped shape this work.

\bibliographystyle{acm}
\bibliography{taco}

\end{document}